\documentclass[
12pt,
3p,
review,
]{elsarticle}
\usepackage{lineno,hyperref}
\modulolinenumbers[1]
\usepackage{graphicx,graphics}
\usepackage{placeins}
\usepackage{amsmath}
\usepackage{amssymb}
\usepackage{mathtools}
\usepackage{array}
\usepackage{lineno,hyperref}
\usepackage{color}
\usepackage{soul}
\usepackage{dcolumn}
\usepackage{bm}
\usepackage{todonotes}
\usepackage{onimage}
\usepackage[utf8]{inputenc}
\usepackage[english]{babel}
\usepackage{libertine}
\usepackage[caption=false]{subfig}
\usepackage{pgfplots}
\usepackage{booktabs}
\usepackage{upgreek}

\usepgfplotslibrary{polar}
\pgfplotsset{compat=1.9}
\usetikzlibrary{calc}

\usepackage{layouts}

\modulolinenumbers[5]

\renewcommand{\l}{\mathopen{}\mathclose\bgroup\left}
\renewcommand{\r}{\aftergroup\egroup\right}

\newcommand{\vphi}{\mbox{{\bm{$\phi$}}}}

\newcommand{\vc}{\mbox{{\bm{$c$}}}}

\newcommand{\vv}[1]{\boldsymbol{#1}}
\newcommand{\n}{\vv{\nabla}}

\newcommand{\diff} [1]{\mathrm{d}{#1}} 
\newcommand{\phia}{\phi_{\alpha}}
\newcommand{\phib}{\phi_{\beta}}

\let\originaleps=\epsilon
\let\epsilon=\varepsilon
\let\varepsilon=\originaleps
\usepackage{stmaryrd}

\definecolor{mydark_blue}{RGB}{0, 0, 139}
\definecolor{myblue}{RGB}{0, 0, 255}
\definecolor{mycyan}{RGB}{0, 255, 255}  
\definecolor{mygreen}{RGB}{0, 255, 0}
\definecolor{myyellow}{RGB}{255, 255, 0}
\definecolor{myred}{RGB}{255, 0, 0}
\definecolor{mydark_red}{RGB}{139, 0, 0}
\definecolor{myblack}{RGB}{0, 0, 0}

\definecolor{BRY_1}{RGB}{  0,  0,255}
\definecolor{BRY_2}{RGB}{127,  0,127}
\definecolor{BRY_3}{RGB}{255,  0,  0}
\definecolor{BRY_4}{RGB}{255,127,  0}
\definecolor{BRY_5}{RGB}{255,255, 85}

\journal{arXiv.org}

\begin{document}

\begin{frontmatter}

\title{Data-driven analysis of grain-growth kinetics in duplex and triplex systems}

\author[mymainaddress1]{P G Kubendran Amos\corref{mycorrespondingauthor}\fnref{fn1}}
\cortext[mycorrespondingauthor]{Prince Gideon Kubendran Amos}
\ead{prince@nitt.edu}

\author[mymainaddress1]{Ramanathan Perumal\fnref{fn1}}
\author[mymainaddress2]{Arnd Koeppe}
\author[mymainaddress2,mysecondaryaddress]{Britta Nestler}

\fntext[fn1]{The authors contributed equally.}

\address[mymainaddress1]{Department of Metallurgical and Materials Engineering,\\ National Institute of Technology, Tiruchirappalli 620015, Tamil Nadu, India}
\address[mymainaddress2]{Institute of Applied Materials (IAM-CMS), Karlsruhe Institute of Technology (KIT),\\
Strasse am Forum 7, 76131 Karlsruhe, Germany}
\address[mysecondaryaddress]{Institute of Digital Materials Science (IDM), Karlsruhe University of Applied Sciences,\\
Moltkestr. 30, 76133 Karlsruhe, Germany}

\begin{abstract} 

Grain growth in multiphase polycrystalline systems are conventionally represented, and discussed, by considering the temporal change in the average radius of the constituent phase-grains, and overall microstructure, separately. 
Despite this elucidation of grain growth, a convincing understanding on how the evolution kinetics of  an individual phase-grains relates to the growth rate of entire microstructure is yet to be achieved. 
Therefore, in this work, comfortably comprehensible statistical tools are employed to realise the influence of the evolving phase-grains on the overall grain-growth kinetics of multiphase systems, both duplex and triplex, with varying phase-fractions. 
A dataset capable of rendering statistically significant, and accessible, information is built by modelling, and analysing, grain growth in twenty different polycrystalline systems, comprising of fourteen triplex and five duplex microstructures with varying phase-fractions, and one homogeneous single-phase system, in multiphase-field framework. 
Statistical analyses of these wide-range of microstructures unravel that, during isotropic grain-growth, the overall evolution kinetics of the duplex and triplex systems are principally governed by the growth-rate of the major-phase grains, irrespective of the phase-fractions. 
In other words, even though introduction of a second phase in a single-phase polycrystalline system considerably reduces the grain-growth rate of the resulting duplex microstructure, the corresponding kinetics of the evolution is predominantly dictated by the growth rate of the major-phase grains with relatively large volume-fraction.  
Furthermore, through multiple non-linear regression, it is realised that with increase in the volume-fraction of the major-phase grains, the overall isotropic grain-growth rate exhibited by duplex and triplex microstructures increase in accordance with a definite exponential relation. 
Additionally, the present investigation unravels that, while the evolution of individual phase-grains during grain growth in triplex microstructures exhibit a level of interdependency, the least interaction is observed between the grains of minor-phases with relatively low volume-fraction. 

\end{abstract}


\end{frontmatter}

\linenumbers

\section{Introduction}

Microstructure of materials are meticulously engineered to cater the ever progressing and more demanding needs. 
Properties, once thought to be irreconcilable, are increasingly combined through appropriate combination of phases in the microstructures~\cite{jacques2001developments,huang2017multiphase,kumar2020carbon}.
Even though microstructures characterised by more than one chemically-distinct phases have been around for a considerable period of time, the distribution of the constituent phases engenders a unique category of multiphase materials. 
For instance, in polycrystalline pearlitic steel, the combination of ferrite and cementite, in a specific phase-fraction, is observed in all grains, despite being separated by the interfaces (grain boundaries)~\cite{zhang2011microstructure,hwang2019microstructure}.
Such grains can be treated as chemically homogeneous, despite the presence of two distinct phases. 
However, on the other hand, there are specialised steels, wherein the grains are not chemically homogeneous but are exclusively associated with one of the constituent phases~\cite{liljas2008development}. 
In other words, instead of martensite and ferrite co-existing in all grains of the polycrystalline microstructure, individual grains assume a phase, thereby establishing a chemical inhomogeneity in the system~\cite{armas2002mechanical}.
Polycrystalline materials, characterised by these chemically-distinct grains, are qualified contextually as multiphase, and depending on number of constituent phases, these materials are referred to duplex, triplex and such. 

Besides steel, multiphase microstructures are established in wide-range of materials to achieve desired combination of properties~\cite{filip2003effect,gollapudi2011microstructure}.  
The extensive applicability of two- and three-phase titanium alloys is primarily due to properties which is a direct consequence of their multiphase microstructure. 
The prevalence of multiphase polycrystalline arrangement in high-entropy alloys is a principal reason for their characteristic behaviour~\cite{tang2015tensile,liu2019fatigue}.  
In ceramics, mechanical properties including fracture toughness are noticeably enhanced by two-phase microstructure.
Duplex system of alumina and silicon carbide is a prime example of such behaviour-enhanced ceramics~\cite{jang1995effect,lutz1991k}. 
Moreover, it has been reported that the introduction of an additional phase, which institutes a triplex microstructure, exacts a more preferred response from ceramics to an imposed mechanical conditions~\cite{neuman2017high,feng2020effect}. 
Similar favourable effects of multiphase microstructure are observed in composite materials as well~\cite{do2008microstructure}. 

In addition to the characteristic feature of the phases associated with the multiphase materials, the emerging properties are also considerably influenced by the volume fraction of the constituent phases, $i.e$ phase-fraction, and average grain size. 
Consequently, to achieve desired behaviour, multiphase materials with varying phase-fraction ranging from minimal minor-phase (close to 1\%) to equi-fraction (50-50 in duplex), and appropriate grain sizes, are fabricated~\cite{sternitzke1997structural,fan1997computer}. 
While material-specific processing techniques are employed for introducing the necessary phase-fraction, the required grain sizes are obtained through a rather well-known microstructural transformation called grain growth~\cite{yu2021high}. 
The mechanism of grain growth in multiphase systems is significantly different single-phase polycrystalline microstructure.
In polycrystalline materials with chemically-homogeneous grains, the local diffusion of atoms across the grain boundaries, which consequently leads to its migration, ultimately governs the grain growth. 
In other words, in single-phase systems, the grain growth is primarily dictated by the movement of the interface (or) grain boundaries. 
However, in multiphase microstructures, the association of grains to a specific constituent phase adds an additional facet to the conventional grain growth. 
Besides reducing the number of grains, grain growth in multiphase systems conserves the characteristic volume-fraction of the phases. 
Therefore, the curvature-driven transformation, which minimises the overall grain-boundary energy, in multiphase materials is, in principle, a combination of coarsening and grain growth~\cite{cahn1991stability}. 
Typifying features of both coarsening and grain growth, which respectively are preservation of volume fraction and growth of larger grains at the expense of smaller ones, are simultaneously observed in multiphase materials. 
For this reason, the energy-minimising curvature-driven transformation in multiphase systems is at times referred to as concurrent grain growth and coarsening. 
Correspondingly, as opposed to interface migration, the long-range diffusion of atoms dictates grain growth in multiphase polycrystalline systems~\cite{fan1997diffusion}. 
Regardless of the mechanism, owing to the influence of the grain size on the properties rendered by the materials, grain growth in multiphase systems have been extensively analysed~\cite{guo2009microstructural,liu2015synergetic}. 
Moreover, it is conceivable, and indeed reported that, a change in the average grain-size of a multiphase material, while employed in an application, results in undesired behaviour. 
In other words, besides the direct effect of grain-size on the properties of multiphase materials, its life in an application is dictated by the rate at which the grains grow.  
Solid-oxide fuels with triplex microstructure are prime examples of this relation between the growth kinetics and life of a material~\cite{lei2017phase}. 
Considering this influence of grain size, investigations have been geared towards understanding the kinetics of grain growth in multiphase polycrystalline system.  

Experimental techniques generally pose definite practical difficulties when adopted for comprehensive analyses of grain growth.
Therefore, theoretical approaches have long since been adopted to complement, and extend, the existing understanding~\cite{saito1992monte,kawasaki1989vertex,anderson1984computer}.
These theoretical studies, particularly one involves multiphase-field models, have been offering critical insights on mechanism and kinetics of grain growth in multiphase system, which are consistent with the experimental observation~\cite{krill2002computer,perumal2017phase,mckenna2014grain}. 
Existing investigations, and resulting understanding, of grain-growth kinetics in multiphase polycrystalline systems can broadly be categorised into two. 
One relates material-specific parameters, including diffusivities and grain-boundary energy anisotropies, to the rate of grain growth~\cite{ravash2014three,ravash2017three2}, while the other focuses on the effect of microstructural features like phase-fractions~\cite{fan1997computer,yadav2016effect}.
While the material-specific studies are, in principle, inherently  bound by the choice parameter(s), the analyses involving phase-fraction offer more generalised insights that are relevant to a wide-range of multiphase polycrystalline systems. 

In-keeping with the change in grain-growth mechanism, the investigations focusing on microstructural features unravel that the introduction of second phases significantly reduces the kinetics of overall evolution. 
Moreover, it is realised that the expression capturing the temporal change in the average radius of the polycrystalline system reflects coarsening kinetics, $\bar{R}^n\propto t$ with $n=3$, irrespective of the volume-fraction of the second phase~\cite{fan1997diffusion,ohnuma1999computer}. 
Even though the major and minor phase(s) - categorised based on volume-fraction- , and the entire polycrystalline system as a whole, evolve at a rate that complies with the coarsening power-law, the kinetic-coefficient ($k$) varies with the  phases, particularly their volumes~\cite{fan1997computer}.
For instances, in a duplex microstructure characterised by unequal volume-fraction of constituent phases, the grains of the major phase, with higher phase-fraction, grow at the noticeably faster rate when compared to the minor phase. 
This disparity in the kinetics of the grain growth between the phases is indicated by the difference in the kinetic-coefficient ($k$).
The relatively sluggish growth of the minor phase(s) is attributed to the increased distance between the respective chemically-similar grains, and complex diffusion-pathways that facilitate the growth, when compared to the proximity of the major phases. 
Furthermore,  attempts have been quantify the effect of volume-fraction on the kinetic-coefficients of major- and minor-phases by comparing the evolution of the microstructures with varying phase-fractions~\cite{yadav2016effect}. 
Despite these apparent advancements, there continues to exist certain aspects to the grain-growth kinetics that prevent a comprehensive understanding of the evolution of multiphase polycrystalline systems. 

In addition to the characteristic nature, and volume-fraction, of the phases, the properties of multiphase systems depend on the \textit{overall grain-size} of the polycrystalline microstructure.
However, studies aimed at explicating the grain-growth rate in multiphase materials primarily report, and discuss, on the difference in the kinetics exhibited by the major- and minor-phases, while largely overlooking the evolution of the entire microstructure~\cite{fan1997topological,jang1995effect}. 
Moreover, the collective influence of the growth rate of major- and minor-phases on the temporal change in overall average size of the grains is yet to be convincingly understood. 
Besides, in three (or more) phase systems, the interdependency between the grain-growth rate of different phases have not been thoughtfully considered so far. 
Amongst others, in the present work, attempts are made to address the aforementioned questions on the grain-growth kinetics of the overall multiphase microstructure through appropriate statistical analysis.

\section{Devising \lq multidimensional\rq \thinspace dataset}

Persuasive understanding on how the evolution of grains of an individual phase relates to the overall grain growth in a multiphase system, can hardly be gained from analysing of a few microstructures.  
To that end, a \lq multidimensional\rq \thinspace dataset is developed by modelling grain growth exhibited by duplex and triplex microstructures in the multiphase-field framework~\cite{perumal2018phase,amos2019understanding}. 
Thermodynamical underpinnings of the adopted model is relatively well-established, and is often acclaimed to be a computationally-efficient alternate~\cite{amos2020grand,amos2020distinguishing}. 
Even though the complete formulation of the present approach~\cite{perumal2019concurrent,perumal2020quadrijunctions}, its ability to incorporate different modes of mass transfer, and recover sharp-interface solutions is exhaustively discussed elsewhere ~\cite{amos2020multiphase,hoffrogge2021multiphase}, a brief outline focusing on the key aspects in rendered here. 

\subsection{Multicomponent multiphase-field model}

Multiphase-field approach is characterised by the introduction of scalar variable(s) which distinguishes the different phases in the system. 
Conventionally, while modelling polycrystalline microstructures, these variables, called phase-field ($\phi(\vv{x},t)$), are employed to differentiate various grains, and their spatio-temporal evolution translates to grain growth. 
Across the grain boundary separating two grains, phase-field assumes a value of $\phi=0$ in one grain, while the other is realised by a constant non-zero value, which is generally $\phi=1$. 
The diffuse region, wherein the value of the phase-field gradually varies, $\phi(\vv{x},t)\in(0,1)$, describes the grain boundary.  
Given that, in most cases, the grains in the polycrystalline systems are chemically homogeneous, phase-field typifying the various grains are introduced as a tuple, which is expressed as 
\begin{align}\label{eq:mp1a}
\vphi=\left \{\phi_{1},\phi_{2},\dots,\phi_{N}\right \},
\end{align}
where $N$ denotes the total number of grains in the system. 
However, since the present approach models grain growth in multiphase systems comprising of $N$ phases, wherein numerous grains are associated each phase, the corresponding phase-field is extended and written as
\begin{align}\label{eq:mp2}
\vphi=\Big\{ \underbrace{ \{ \phia^{1},\phia^{2}\dots\phia^{q_{\alpha}} \}}_{\vv{\phia}}, \underbrace{ \{ \phib^{1},\phib^{2}\dots\phib^{q_{\beta}} \}}_{\vv{\phib}}\dots \underbrace{ \{ \phi_{N}^{1},\phi_{N}^{2}\dots\phi_{N}^{q_{N}} \}}_{\vv{\phi_{N}}} \Big\}.
\end{align}
Moreover, in this formulation, number of grains sharing a given phase is represented by $q_i$ where $i\in\{\alpha,\beta,\dots,N\}$.
By assigning appropriate concentration, grains of a given phase are distinguished from the rest, $\{\vphi(\vc)=\vphi_\alpha(\vc) | \vc=\vv{c}_\alpha\}$.  
Similar to phase-field, in order to encompass the $K$-different chemical components, the concentration is introduced as a tuple.
Correspondingly, concentration of a random grain $m$ associated with phase-$\alpha$ reads
\begin{align}\label{eq:conc1}
\vc^{\alpha}_{m}=\left\{c^{\alpha}_{m:i},c^{\alpha}_{m:j},\dots,c^{\alpha}_{m:K}\right\}.
\end{align}
The homogeneity in the chemical composition of the grains associated with a given phase, say $\alpha$, yields
\begin{align}\label{eq:conc2}
\vc^{\alpha}_{1}=\dots=\vc^{\alpha}_{m}=\dots=\vc^{\alpha}_{q_{\alpha}}\equiv\vc^{\alpha}=\left\{c^{\alpha}_{i},c^{\alpha}_{j},\dots,c^{\alpha}_{k}\right\}.
\end{align}
To ensure that the volume-fraction of the phases remains unaltered despite the evolution, equilibrium composition is assigned to the corresponding phases.
In other words, since the current approach attempts to model grain growth in multiphase system, $\vc^{\alpha}$ represents a tuple of equilibrium composition that characterises phase-$\alpha$.

Following the conventional framework, the overall energy-density of the multiphase polycrystalline system is formulated as the combination interface (grain boundaries) and bulk (grain) contribution~\cite{provatas2011phase}. 
Correspondingly, by incorporating the appropriate phase-field and concentration, the energy-density of the system comprising of $N$-phases and $K$ chemical components is written as 
\begin{align}\label{eq:functional1}
\cal{F}(\vphi,\n \vphi,\vc) & =\cal{F}_{\text{int}}(\vphi,\n \vphi)+\cal{F}_{\text{bulk}}(\vphi,\vc) \\ \nonumber
&=\int_{V}f_{\text{int}}(\vphi,\n \vphi)+f_{\text{bulk}}(\vphi,\vc)\diff V,
\end{align}
where $f_{\text{bulk}}(\vphi,\vc)$ and $f_{\text{int}}(\vphi,\n \vphi)$ are the respective energy-contribution of grain and grain boundary with $V$ representing the volume. 
The interface contribution, $f_{\text{int}}(\vphi,\n \vphi)$, akin to most multiphase-field techniques, comprises of a gradient- and potential-energy term. 
Grain-boundary energy densities, and corresponding anisotropies, are introduced to the system through the interface-energy contribution~\cite{tschukin2017concepts}. 
While multi-well potentials are generally employed to penalise phase-field, and ensure its bounds, in the present work, an obstacle-type potential operating in combination with Gibbs simplex is involved~\cite{amos2018phase}. 
Furthermore, the energy contributions from the grains, $f_{\text{bulk}}(\vphi,\vc)$, which is reasonably assumed to be insignificant while modelling grain growth in single-phase system, is formulated as the interpolation of the energy contribution of the individual grains, $f(\vphi,\vc)=\sum_{\alpha}^{N}\sum_{m}^{q_{\alpha}}f^{\alpha}_{m}(\vc^{\alpha})h(\phia^{m})$.
Given that the contribution of the individual grains is dictated by the characteristic equilibrium-composition of the associated phases, the volume of the phases during grain growth is preserved by the energy-density term, $f_{\text{bulk}}(\vphi,\vc)$~\cite{amos2018globularization,amos2020limitations}.  

The spatio-temporal evolution of phase-field, which translates to grain growth, is formulated by considering phenomenological minimisation of the overall energy-density of the system.
Correspondingly, the evolution of a random grain $m$, which is associated with phase-$\alpha$, is dictated by
\begin{align}\label{eq:ph_evo1}
\tau\epsilon\frac{\partial \phi_{\alpha}^{m}}{\partial t} & = -\frac{\partial \cal{F}(\vphi,\n \vphi,\vc)}{\partial \phi_{\alpha}^{m}}\\ \nonumber
&=\epsilon\left[ \n \cdot \frac{\partial a(\vphi, \n \vphi)}{\partial \n \phi_{\alpha}^{m}} - \frac{\partial a(\vphi, \n \vphi)}{\partial \phi_{\alpha}^{m}} \right] - \frac{1}{\epsilon} \left[ \frac{\partial w(\vphi)}{\partial \phia^{m}} \right] - \left[ \frac{f^{\alpha}_{m}(\vc^{\alpha},\phia^{m})}{\partial \phia^{m}} \right] - \Lambda,
\end{align}
where the Lagrange multiplier $\Lambda$ is introduced to ensure that the summation of phase-fields at any point in the system is $1$.
Moreover, in the above evolution equation, while $a(\vphi, \n \vphi)$ and $w(\vphi)$ correspond to the gradient and potential energy terms, the parameter dictating interface-width, and its stability during the migration, are denoted by $\epsilon$ and $\tau$, respectively.

In multiphase-field models, wherein the energy contribution of the bulk phases are described based on the dependent-concentration, the corresponding driving-force, which emerges from $\frac{f^{\alpha}_{m}(\vc^{\alpha},\phia^{m})}{\partial \phia^{m}}$ in Eqn.~\eqref{eq:ph_evo1}, and dictates the evolution of phase-field, can be viewed as the difference in the Legendre transform of the free-energy densities.
This understanding forms the basis of the grand-potential approach, and when consistently extended assumes chemical potential as the continuous and dynamic variable replacing phase-dependent concentration~\cite{plapp2011unified}. 
Given its computational efficiency, this approach is adopted in the present work, and the driving-force dictating phase-field evolution is formulated by treating chemical potential as the dynamic variable. 
The temporal evolution of the chemical potential, which principally governs the bulk driving-force in phase-field evolution, is written as
\begin{align}\label{chempot_ev}
\frac{\partial \mu_{i}}{\partial t}=\left \{ \n\cdot\left[ \sum_{j=1}^{K-1} \vv{M}(\vphi) \n \mu_{j}\right ] - \sum_{\alpha}^{N} \sum_{m}^{q} c_{i}^{\alpha}\frac{\partial \phia^{m}}{\partial t} \right \} \left [ \sum_{\alpha}^{N} \sum_{m}^{q} h(\phia^{m}) \frac{\partial c_{m:i}^{\alpha}}{\partial \mu_{j}} \right ]_{ij}^{-1},
\end{align}  
where $\mu_{i}$ denotes the continuous chemical-potential of component-$i$. 
The mobility of the migrating elements, in the multicomponent setup, is dictated by matrix $\vv{M}(\vphi)$ which also facilitates the incorporation surface diffusion~\cite{amos2020multiphase,hoffrogge2021multiphase}. 
The phase-dependent concentration of component-$i$ in random grain $m$ belonging to phase-$\alpha$, and the corresponding interpolation function, are represented by $c_{m:i}^{\alpha}$ and $h(\phia^{m})$, respectively.
The evolution of the different microstructures with varying phase-fractions, in this work, are modelled by solving the Eqns.~\eqref{eq:ph_evo1} and ~\eqref{chempot_ev}.

\subsection{Simulation setup}

Existing studies unravel that, unlike Zener pinning~\cite{fan1998numerical}, the overall trend in grain-growth kinetics exhibited by the individual phases, and entire microstructure, of the multiphase system are largely independent of the dimensionality of the simulation domain~\cite{fan1997computer,poulsen2013three,yadav2016effect}.
In other words, in both two- and three-dimensional setup, similar disparity in the evolution kinetics of  major-, minor- and equifraction phases, in relation to overall growth rate, has been reported. 
Therefore, in the present work, the grain growth in various multiphase-polycrystalline systems are modelled in two-dimensional framework. 
Moreover, irrespective of the variation in phase-fraction, two-dimensional domains of similar configurations are adopted for all numerical studies.

Two-dimensional domains considered for the current investigations are uniformly discretised into $2048 \times 2048$ cells of identical dimension, $\Delta$x = $\Delta$y = $5 \times 10^{-7}$m, through the finite-difference scheme.
Polycrystalline microstructures comprising of approximately $10000$ grains are instituted over the discretised domain through Voronoi tessellation. 
These grains are associated to the constituent phases by assigning the characteristic chemical composition.
Since the grains are equiaxed with almost similar size, the required phase-fraction in the microstructure is achieved by relating the appropriate number of randomly distributed grains to the corresponding phases.
In other words, duplex microstructure comprising of 33$\%$ minor phase is devised, in the initial stages, by assigning the respective chemical composition to the one-third of the grains randomly. 

\begin{figure}
    \centering
      \begin{tabular}{@{}c@{}}
      \includegraphics[width=1.0\textwidth]{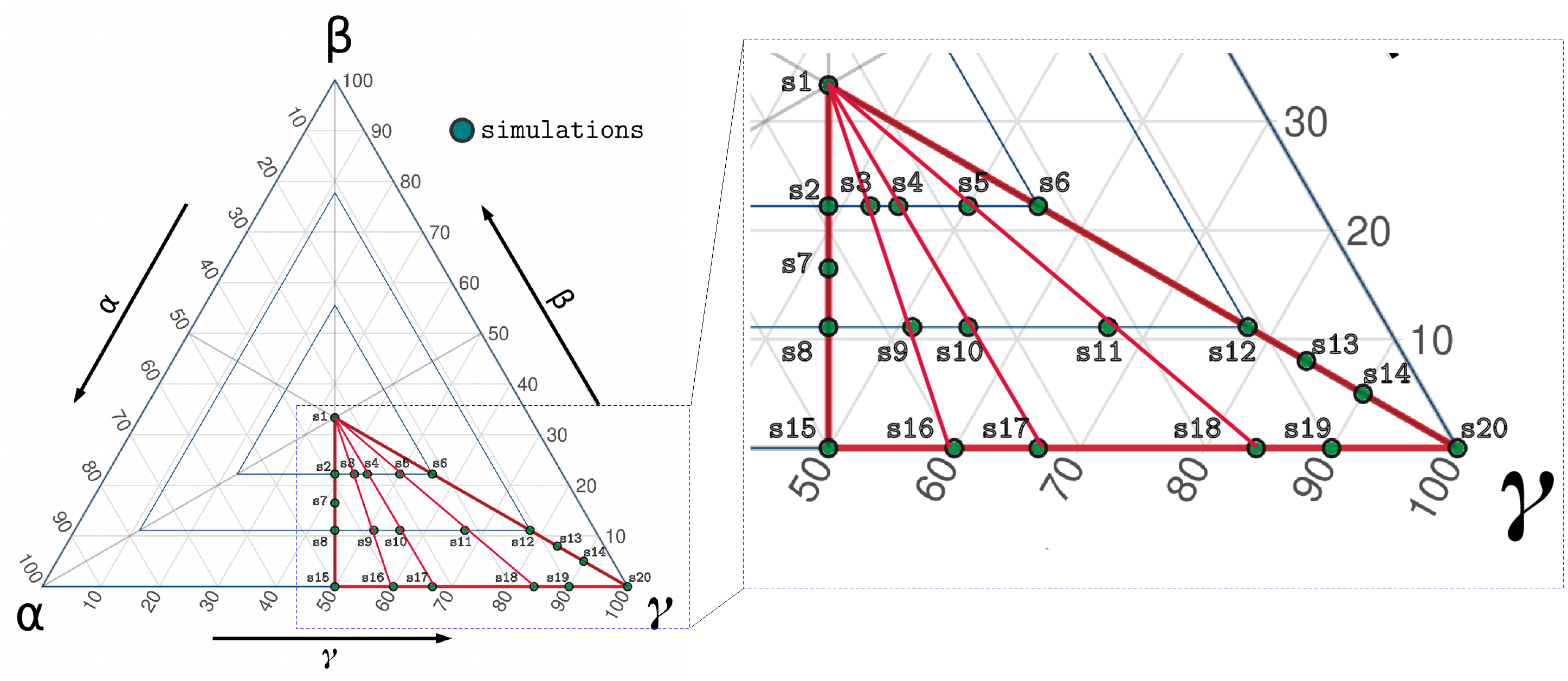}
    \end{tabular}
    \caption{  A simplex, analogous to ternary isotherm, depicting all-possible phase-fractions in duplex and triplex microstructures, along with single-phase systems (vertices). The section that encompasses the varied microstructures considered in this work is distinguished, and each  point referring to a specific system is highlighted and designated as $s1$, $s2$, and such. 
    \label{fig:simplex}}
\end{figure}

Given that the principal focus of this work is to understand the evolution kinetics of  different phases in relation to each other, and to the overall grain-growth rate exhibited by the entire microstructure, a rather straightforward distinction is made between the phases. 
While a binary system with two chemical component, $\tilde{i}$ and $\tilde{j}$, is considered for establishing duplex microstructure, three-phase microstructure is construed in the framework of ternary system with components $\tilde{i}$, $\tilde{j}$ and $\tilde{k}$.
In duplex system, $\alpha$-phase is a $\tilde{i}$-rich phase with equilibrium composition of $c^{\alpha}_{\tilde{i}:eq}=0.9$, and $c^{\alpha}_{\tilde{j}:eq}=0.9$ characterises matrix $\gamma$-phase, wherein both concentrations are expressed in mole fraction.
Concentration of the solvent in $\alpha$- and $\gamma$-phase remains unaltered in the triplex system, while the remnant content is equally partitioned between solutes, $\{\tilde{j},\tilde{k}\}$ and $\{\tilde{i},\tilde{k}\}$, respectively.
The $\beta$-phase, exclusively introduced in the three-phase systems, is characterised by composition $c^{\beta}_{\tilde{i}:eq} = c^{\beta}_{\tilde{j}:eq} = 0.05$.
Since the present investigation is primarily interested in microstructural features like phase-fractions, diffusivities of the components are assumed to be identical, and unit matrix is correspondingly incorporated in the formulation. 
Moreover, the energy-densities of the grain boundaries, irrespective of the chemical-composition of the grains they separate, is treated as isotropic and unity. 
The length scale parameter, $\epsilon$, is appropriately defined such that the diffuse interface is of constant thickness comprising of four cells~\cite{mittnacht2021morphological}. 

The temporal evolution of the dynamic variables, phase-field and chemical potential, that dictate the microstructural changes in the multiphase polycrystalline system, are solved over the homogeneous cells of the two-dimensional domain by forward-marching Euler's scheme. 
In order to ensure that the computational resources are optimally used, the domain is decomposed into smaller segments, and dealt simultaneously, through Message Passing Interface (MPI). 
The complexity of the numerical treatment is reduced by suitably non-dimensionalising the input parameters, and incorporating them as dimensionless values~\cite{amos2020multiphase}.

\subsection{Varying phase-fractions}

\begin{figure}
    \centering
      \begin{tabular}{@{}c@{}}
      \includegraphics[width=1.0\textwidth]{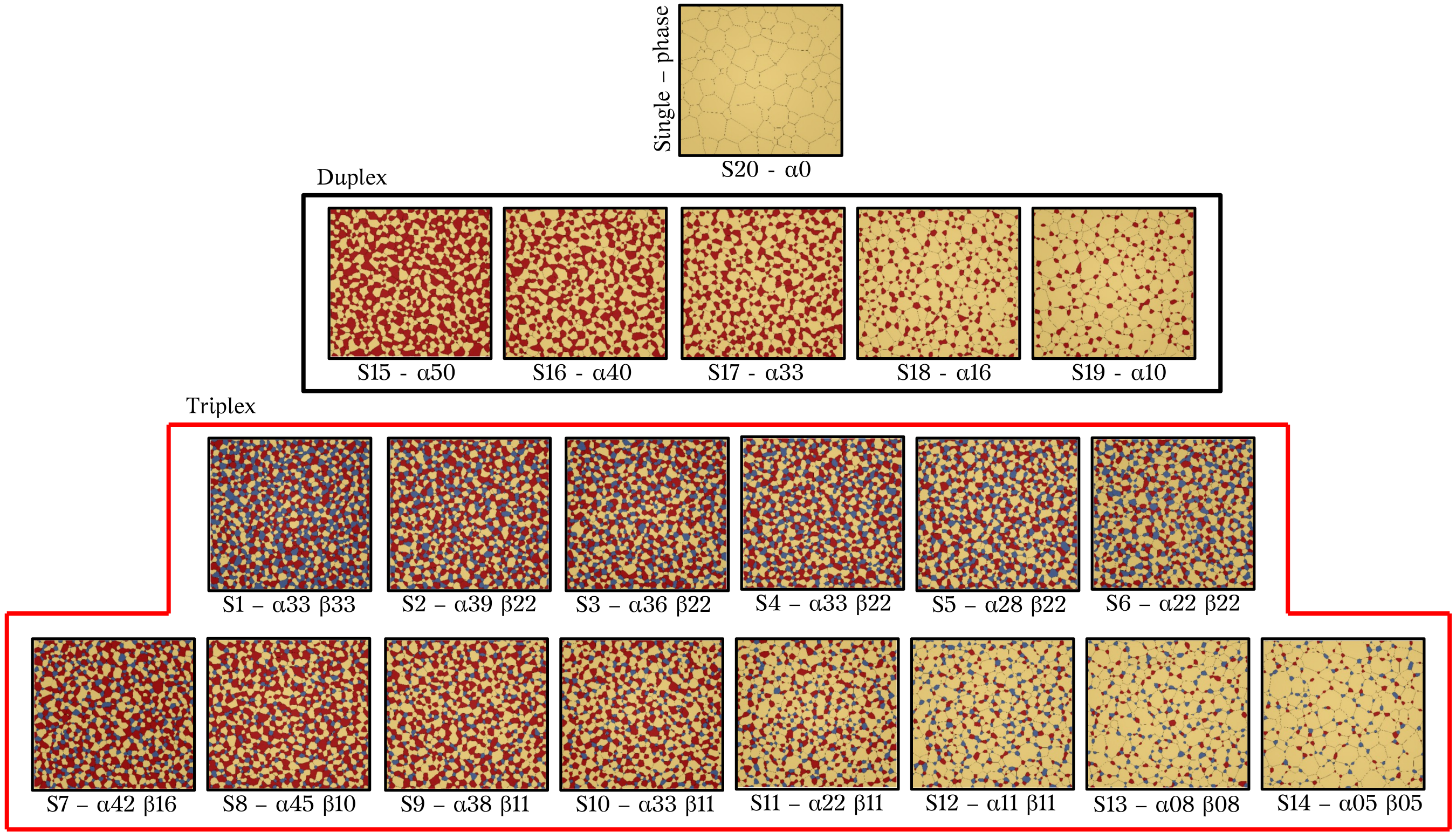}
    \end{tabular}
    \caption{ Microstructure corresponding to each point in the section of the simplex that includes homogeneous systems along with duplex and triplex microstructures with varying volume-fraction of the constituent phases.  
    \label{fig:systems}}
\end{figure}

In applications, depending on the material need, multiphase systems with varying degree of phase-fractions are employed. 
Though investigating every combination of the phase-fraction would be redundant, convincing level of understanding can only be gained by considering relatively increased number of phase-fractions. 
Particularly, since the present study adopts statistical techniques to explicate the kinetic relation between the evolving phases, the accuracy of the outcome depends on the wealth of information (data) available. 
To that end, in this work, grain growth in twenty different systems, which encompasses  one homogeneous, five duplex and fourteen triplex microstructures, is modelled, and \lq multidimensional\rq \thinspace dataset is built by monitoring the temporal evolution of the grains. 

As opposed to random consideration of different volume-fractions of phases, a systematic choice of various phase-fraction is made from a 2-simplex.
The simplex, in its entirety, along with the section focused for the current study is shown in Fig.~\ref{fig:simplex}. 
The points within, and on, the 2-simplex can be interpreted in a manner akin to the ternary isotherm.
Correspondingly, while the three vertices indicate the homogeneous microstructure of phase-$\alpha$, -$\beta$ and -$\gamma$, the duplex microstructures are encapsulated by the edges joining the vertices. 
Any point within the simplex represents triplex system, with phase-fraction dictated by its position.
As illustrated in Fig.~\ref{fig:simplex}, a section of the simplex emanating from the vertex characterising the homogeneous $\gamma$-microstructure is considered for the present analyses. 
This section of the simplex renders a wide-variety of polycrystalline systems ranging from single phase homogeneous to triplex with equifraction of constituent phases. 
Moreover, owing to the configuration of the section in Fig.~\ref{fig:simplex}, $\gamma$ -phase acts as the matrix for the duplex and triplex microstructures with unequal volume-fractions of phases. 
Multiphase microstructures corresponding to the different points of the simplex-section is collectively illustrated in Fig.~\ref{fig:systems}. 
Though the volume-fraction of the minor-phases can be as low as 5$\%$, the grains associated with these phases hardly occupy a position on the grain boundary. 
In other words, despite the low volume-fraction and reduced size, the grains of the minor phases seldom render an influence analogous to the particles in Zener pinning. 
Moreover, the reduced size of the grains associated with minor phases, in the initial stages of the grain growth, is consistent with experimental observations~\cite{liu2015synergetic,ritasalo2013microstructural,praveen2016exceptional} and existing theoretical studies~\cite{fan1997computer,yadav2016effect}. 

\section{Results and discussion}

\begin{figure}
    \centering
      \begin{tabular}{@{}c@{}}
      \includegraphics[width=0.65\textwidth]{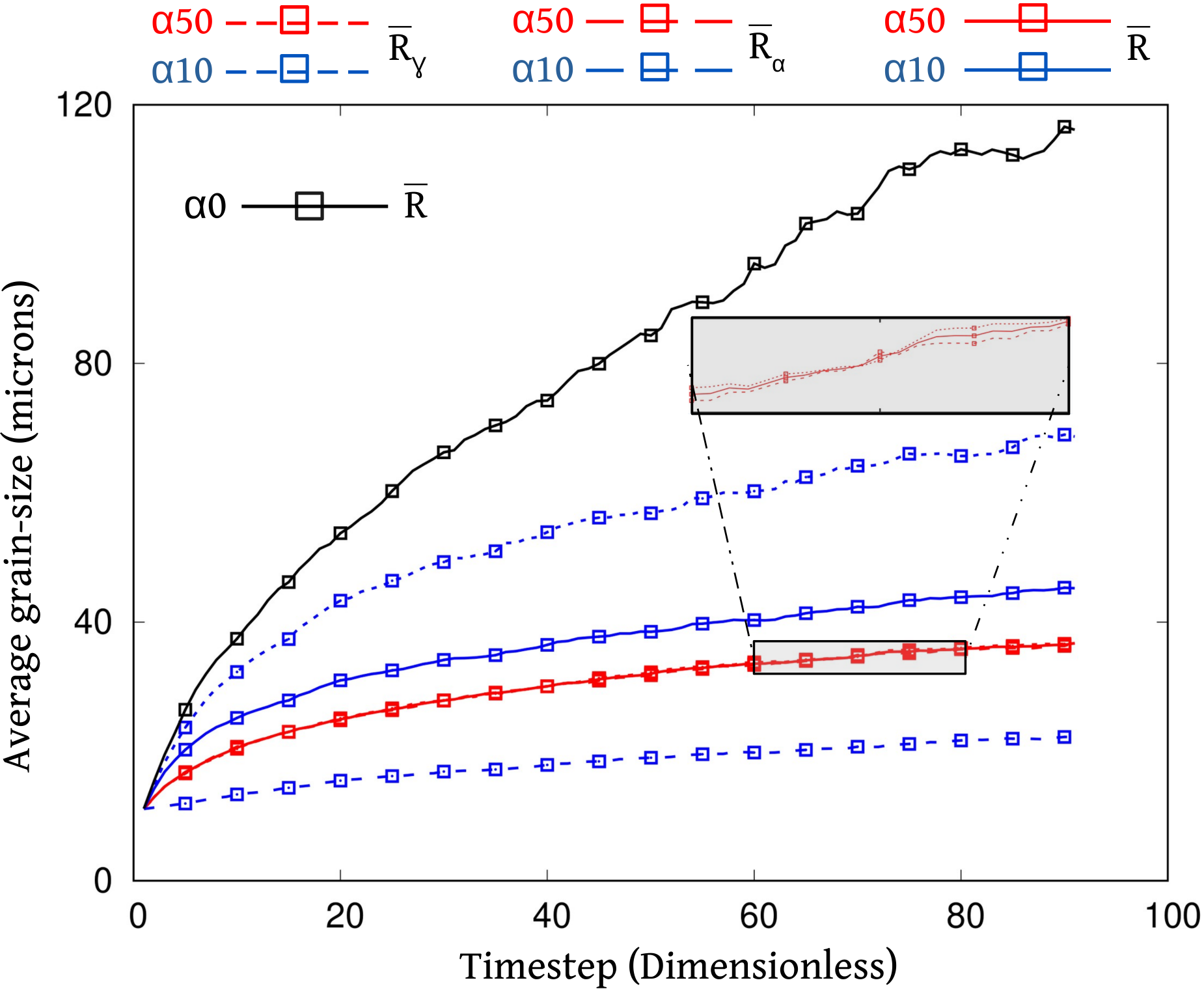}
    \end{tabular}
    \caption{  Temporal change in the average radius of the phase-$\alpha$ and $-\gamma$ grains, $\bar{R}_\alpha(t)$ and $\bar{R}_\gamma(t)$, and the entire microstructure,  $\bar{R}(t)$, for homogeneous, and duplex systems characterised by $50\%$ and $10\%$ of phase-$\alpha$. 
    \label{fig:duplex_validation}}
\end{figure}

By monitoring the grain growth exhibited by the homogeneous and multiphase systems, a \lq multidimensional\rq \thinspace dataset is devised which essentially comprises of temporal change in the average radius of the individual phase-associated grains, and the entire microstructure as a whole.
(A dataset can truly be multidimensional only when its dependent variable is governed by more than one independent variable. 
Considering that explicating the multivariate nature of the overall grain-growth kinetics in multiphase system is the ultimate outcome of the current study, the term \lq multidimensional\rq \thinspace is written within quotes.)
This dataset is analysed through comfortably realised statistical techniques to unravel the effect of the individual phases on the evolution kinetics of the entire system.  
Since the present study focuses primarily on the kinetics of the grain growth, the topological changes and the distribution of the grains are largely overlooked. 

\subsection{Duplex microstructures}

Duplex microstructure comprises of two distinct phases, and is characterised by grains associated with one of these constituent phases, $\alpha$ and $\gamma$. 
Despite the inhomogeneity in the concentration distribution, in duplex microstructures, given that the grain growth occurs in a continuum, the temporal evolution of one phase, and its corresponding kinetics, is inherently coupled with the other. 
This is illustrated and discussed in Appendix.
In other words, the growth rate exhibited by the duplex microstructures with varying phase-fractions can be convincingly expressed by considering kinetics of the only one of the evolving phases. 
Therefore, the aim of the present investigation in duplex microstructure reduces to identifying which of the phases, major or minor, principally governs the kinetics of overall grain-growth. 

\subsubsection{Comparison with single phase microstructure}

Before proceeding to realise the degree of influence rendered by the different phases on the overall growth-kinetics exhibited by the duplex microstructure, rather straightforward investigations are pursued to verify the outcomes of the present approach in relation to the existing reports~\cite{fan1997computer,fan1997diffusion}. 
Even though,  the outcomes of the modelling technique in relation to the established theories and observations have already been reported elsewhere~\cite{amos2020multiphase}, certain relevant aspects are analysed in a statistical framework here. 

In Fig.~\ref{fig:duplex_validation}, the progressive change in the average radius of homogeneous and two duplex microstructures with time are presented. 
Moreover, the temporal increase in the average radius of the phase-associated grains are monitored, and included in this illustration. 
While $\bar{R}(t)$ represents the average radius of the entire polycrystalline microstructure, the corresponding parameter for the grains of  phase-$\alpha$ and -$\gamma$ in duplex microstructures is respectively denoted by $\bar{R}_\alpha(t)$ and $\bar{R}_\gamma(t)$.
Moreover, the multiphase microstructures in this, and subsequent, discussions are described based on the volume fraction of the minor phase. 
For instance, $\alpha 10$ indicates duplex microstructure with $10\%$ phase-$\alpha$, while equifraction system are denoted $\alpha 50$.

Irrespective of the nature of the microstructure, homogeneous or otherwise, Fig.~\ref{fig:duplex_validation} shows a continual increase in the average radius reflecting the grain growth exhibited by the system. 
With the introduction of a second-phase in the microstructure, a significant decrease is observed in the rate at which the radius increases with time. 
This noticeable change in the kinetics is predominantly due to the change in grain-growth mechanism, which is governed by the long-range diffusion of the chemical components in duplex microstructure.
Moreover, in system with equal volume-fraction of phases, the temporal increase in average radius of the entire microstructure and individual phases are largely identical with marginal deviation. 
On the other hand, in $\alpha 10$, significant disparity is noticed in the rate at which the major-phase grains evolve when compared to the minor-phase. 
Fig.~\ref{fig:duplex_validation} illustrates that the increase in the average radius of the entire duplex microstructure lies in between the growth exhibited by the individual phases.
The difference on the growth kinetics between the phases, exclusively in the duplex microstructure characterised by unequal volume-fraction, is due to the corresponding distribution of the phases. 
Owing to its reduced volume, the grains of the minor phase in $\alpha 10$ microstructure are considerably separated when compared to major-phase grains. 
Therefore, the diffusion path, which the chemical components need to transverse to achieve grain growth, is longer, and more convoluted.
Consequently, the growth rate exhibited by the minor-phase grains is significantly lower the grains of phase-$\gamma$. 
So far, no convincing argument has been made on how the evolution kinetics of the entire duplex microstructure relates to the different growth-rates adhered-to by the major- and minor-phase grains, and to that end, this becomes the primary focus of the current analysis. 

\begin{figure}
    \centering
      \begin{tabular}{@{}c@{}}
      \includegraphics[width=0.85\textwidth]{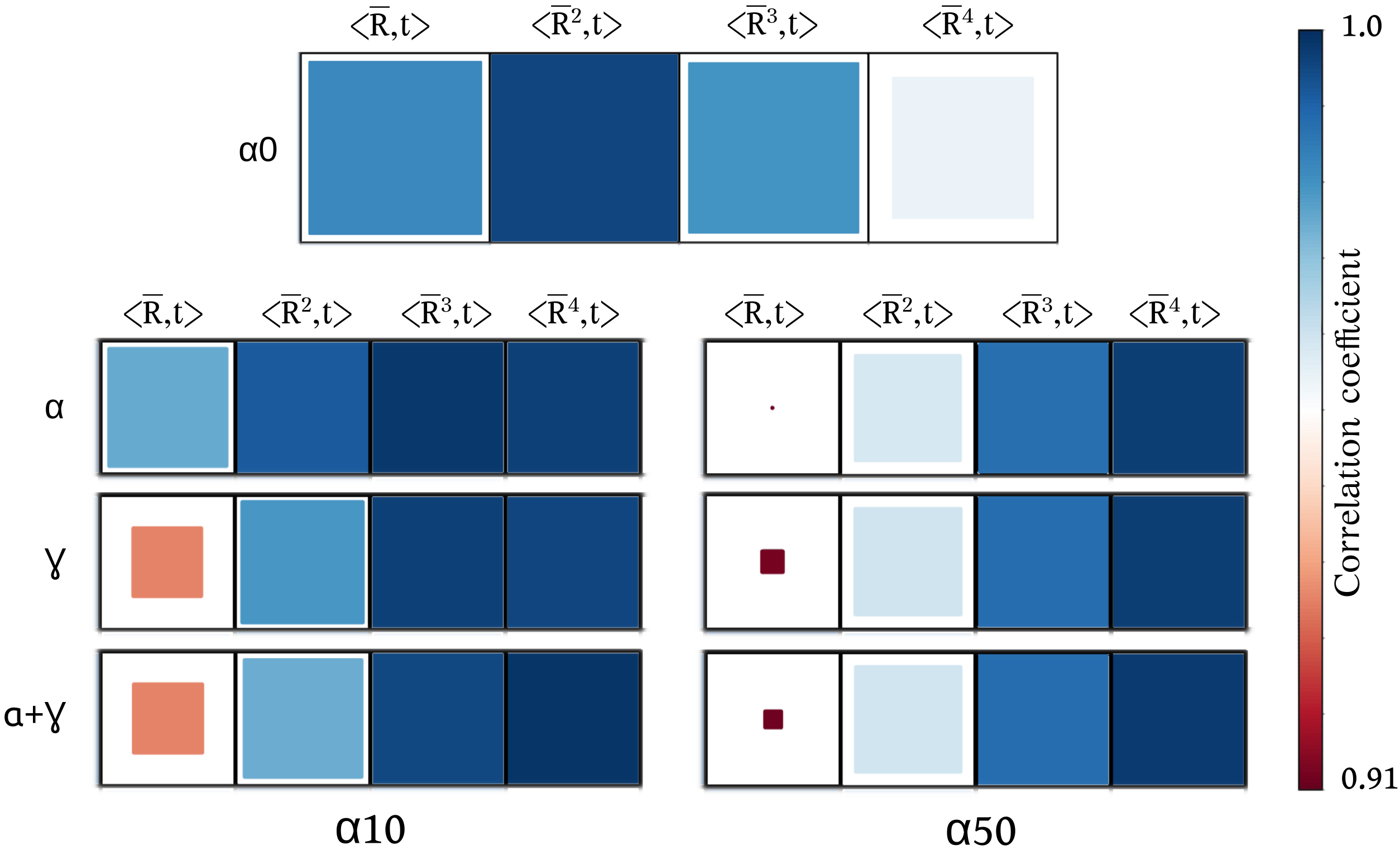}
    \end{tabular}
    \caption{  Correlation coefficient characterising the proportionality between time and the average radius of individual phase-grains and entire microstructure, raised to different exponents,  for homogeneous and duplex systems $\alpha50$ and $\alpha10$.  
    \label{fig:duplex_power}}
\end{figure}

\subsubsection{Grain growth kinetics}

Even though the microstructures illustrated in Fig.~\ref{fig:duplex_validation} render a progressive increase in the average radius with time, owing to the difference in the governing mechanism, the exponent of the power law capturing the growth kinetics vary depending on the nature of the system~\cite{fan1997computer,fan1997diffusion}. 
While the exponent $n=2$ characterises grain growth in homogeneous system, evolution kinetics of the individual phases, and the duplex microstructure as a whole, largely follow relation $\bar{R}^3(t) \propto t$.
In order to affirm that the evolution of the entire microstructure, and its corresponding phases, adhere to the power law,  the temporally varying average radius is raised to different exponents, $n=\{1,2,3,4\}$, and related to time. 
The correlation coefficient (Pearson) characterising the relation between the average radius with the various exponent and time is ascertained, and graphically represented in Fig.~\ref{fig:duplex_power}. 
In this illustration, $<\bar{R}^n,t>$ denotes the correlation coefficient between the average radius raised to order n and time. 
Fig.~\ref{fig:duplex_power} shows that, in homogeneous system ( $\alpha 0$), maximum correlation is observed when $n=2$, thereby indicating that the grain growth in this microstructure adheres to the power law, $\bar{R}^2(t) \propto t$.
Furthermore, correlation coefficient relating the average radius of the individual phases, and entire duplex microstructures, with time, is higher when $n=3$, which implies that the evolution in the multiphase systems complies to the established power law~\cite{fan1997diffusion}. 
Even though it might appear that,  in Fig.~\ref{fig:duplex_power}, for equifraction duplex system ($\alpha 50$) , the maximum correlation is observed in $n=4$.
However, given the marginal difference when compared to $n=3$, such consideration leads to \lq overfitting\rq \thinspace, thus returning to $\bar{R}^3(t) \propto t$ as the statistically sound relation. 

\subsubsection{Ascertaining governing factor}

As shown in Fig.~\ref{fig:duplex_validation}, the change in the average radius of the entire microstructure with time, in duplex systems with unequal volume-fraction of phases, invariably lies  between curves representing the increase in the radius of the individual phase-grains. 
Moreover, existing works unravel that, in radius versus time plot, the relative position of the overall microstructure curve varies with respect to the phase-associated grains curves, as the phase-fraction changes~\cite{fan1997computer,yadav2016effect}.
While the effect of phase-fraction on the evolution of the individual phases are convincingly elucidated, its influences on the growth rate of entire duplex microstructure is yet to be sufficiently addressed. 
In order to identify this influence, more particularly, to answer the resulting question on evolution kinetics of which phase, major or minor, principally effects the growth rate of entire duplex microstructure, each system is statistically analysed.
Statistical programming language R is employed for these, and all other relevant, investigations in this work. 

\begin{figure}
    \centering
      \begin{tabular}{@{}c@{}}
      \includegraphics[width=0.9\textwidth]{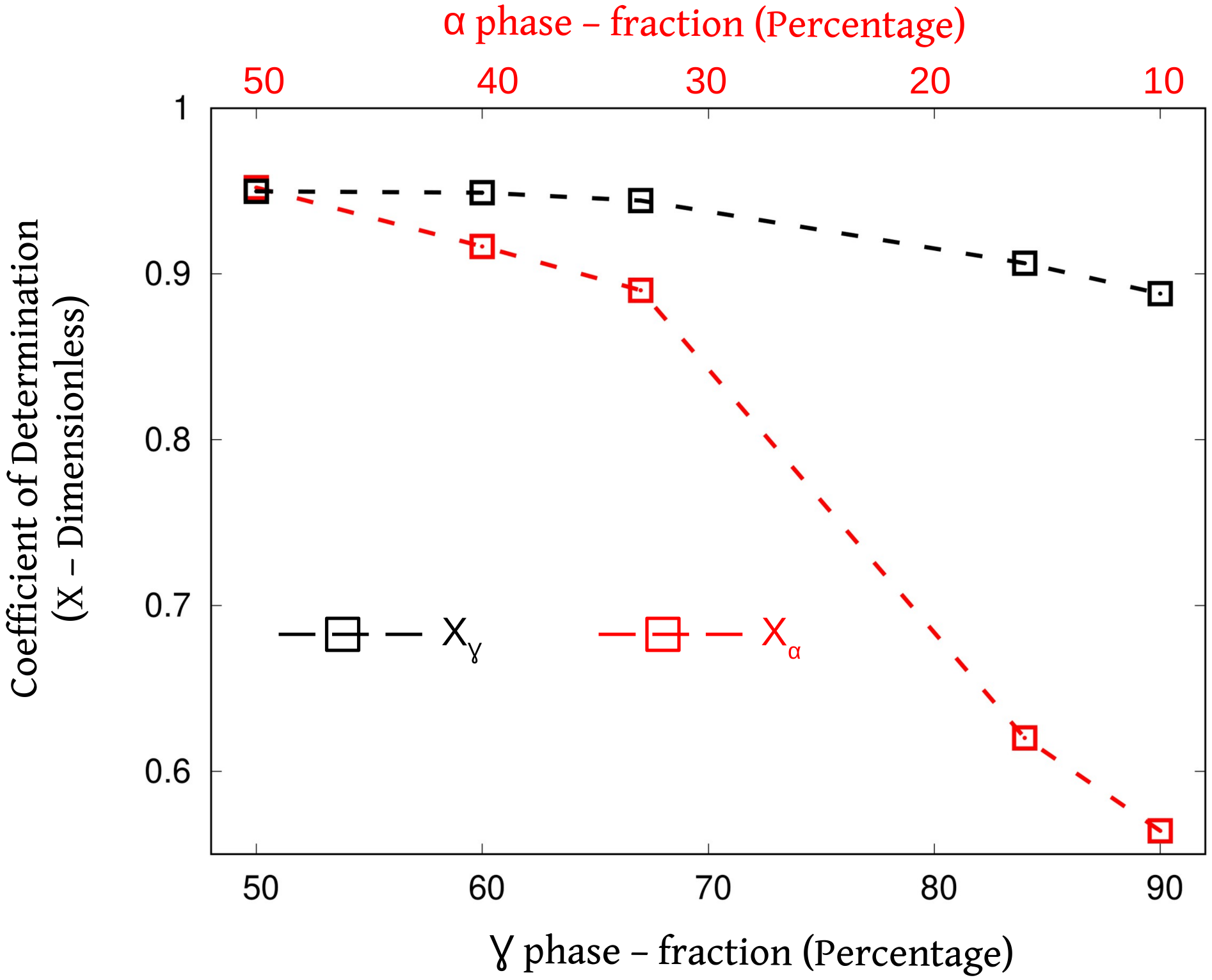}
    \end{tabular}
    \caption{  Coefficient of determination, quantifying the effect of  growth kinetics of individual phase-grains on the overall grain-growth rate of the entire system, is  estimated using Eqn.~\eqref {coeff_deter} for different duplex microstructure with varying phase-fraction.
    \label{fig:rdq_duplex}}
\end{figure}

From the dataset comprising of  temporally varying average radius, the growth rate exhibited by the individual phase-grains ($d\bar{R}_{\alpha}/dt$ or $d\bar{R}_{\gamma}/dt$) and entire microstructure ($d\bar{R}/dt$), at every instance $(t)$, is determined for each duplex system. 
Subsequently, by treating the growth rate of the individual phase-grains, and entire microstructure, as \textit{response} and \textit{predictor} variable, respectively, the corresponding kinetics are related to the each other.
From the emerging relation, the \textit{coefficient of determination}, $\chi$, for the combination of an individual phase-grains and overall microstructure is estimated.
Scatter plots that illustrates the dependency of $d\bar{R}_{\alpha}/dt$ or $d\bar{R}_{\gamma}/dt$ and $d\bar{R}/dt$ are included in \textcolor{red}{Appendix2}.

For a given duplex system, following the conventional description, the coefficient of deterioration considering the growth kinetics of minor phase-$\alpha$ ($d\bar{R}_{\alpha}/dt$) and entire microstructure ($d\bar{R}/dt$) is calculated by 
\begin{align}\label{coeff_deter}
 \chi_{\alpha}=\frac{\chi^{\text{SST}}_{\alpha}-\chi^{\text{SSE}}_{\alpha}}{\chi^{\text{SST}}_{\alpha}},
\end{align}
where $\chi^{\text{SST}}_{\alpha}$ is estimated by treating the instantaneous growth-rate of overall microstructure as univariate parameter, and summing-up the squares of the disparity (error) between the individual values and the mean.  
On the other hand, $\chi^{\text{SSE}}_{\alpha}$ represents the sum of the squared differences between the datapoints and regression line relating the instantaneous kinetics of the $\alpha$-phase grains and overall duplex-microstructure. 
Based on the description of the coefficient of determination in Eqn.~\eqref{coeff_deter}, $\chi_{\alpha}$ can be viewed as a parameter that quantifies the effect of $\alpha$-grains growth-kinetics on the evolution rate of entire microstructure.  
Therefore, in addition to $\chi_{\alpha}$, the corresponding parameter that realises the influence of the major-phase growth-kinetics ($d\bar{R}_{\gamma}/dt$) on the overall evolution rate, $\chi_{\gamma}$ is appropriately determined for all the different duplex microstructures considered in this investigation. 

Coefficients of determination separately quantifying the role of $d\bar{R}_{\alpha}/dt$ and $d\bar{R}_{\gamma}/dt$ in overall growth-rate exhibited by the microstructure, $\chi_{\alpha}$ and $\chi_{\gamma}$, is calculated for different duplex systems with varying phase-fractions and plotted in Fig.~\ref{fig:rdq_duplex}. 
The variation observed in the coefficients of determination, across the different duplex systems, unravels that the effect of individual phase-grains on the overall growth kinetics is primarily dependent on phase-fraction of the microstructure. 
In a duplex system characterised equal volume-fraction of phases, identical coefficients of determination implies that both $\alpha$- and $\gamma$-grains similarly influence the evolution kinetics of the entire microstructure. 
On the other hand, noticeable disparity between $\chi_{\alpha}$ and $\chi_{\gamma}$ is observed in duplex microstructures with varying volume-fraction of constituent phases. 
Moreover, Fig.~\ref{fig:rdq_duplex} shows that this inequality in the coefficients of determination becomes more pronounced with increase in the difference  between the volume-fraction of the phases in duplex microstructure.
While the coefficient of determination pertaining to major phase-$\gamma$ exhibits a relatively marginal change, and continues to remain noticeably greater, $\chi_{\alpha}$ progressively decrease with reduction in the volume-fraction of the corresponding minor-phase grains. 
In other words,  Fig.~\ref{fig:rdq_duplex} unravels that, in duplex systems with unequal volume-fraction of phases, the overall growth rate of the entire microstructure ($d\bar{R}/dt$) is primarily influenced by the evolution kinetics of the major-phase grains ($d\bar{R}_{\gamma}/dt$). 
Furthermore, it is evident from the illustration that the dominance of the major phase in effecting the overall growth kinetics becomes more definite with increase in the corresponding volume-fraction (or decrease the amount of minor phase).

\begin{figure}
    \centering
      \begin{tabular}{@{}c@{}}
      \includegraphics[width=0.9\textwidth]{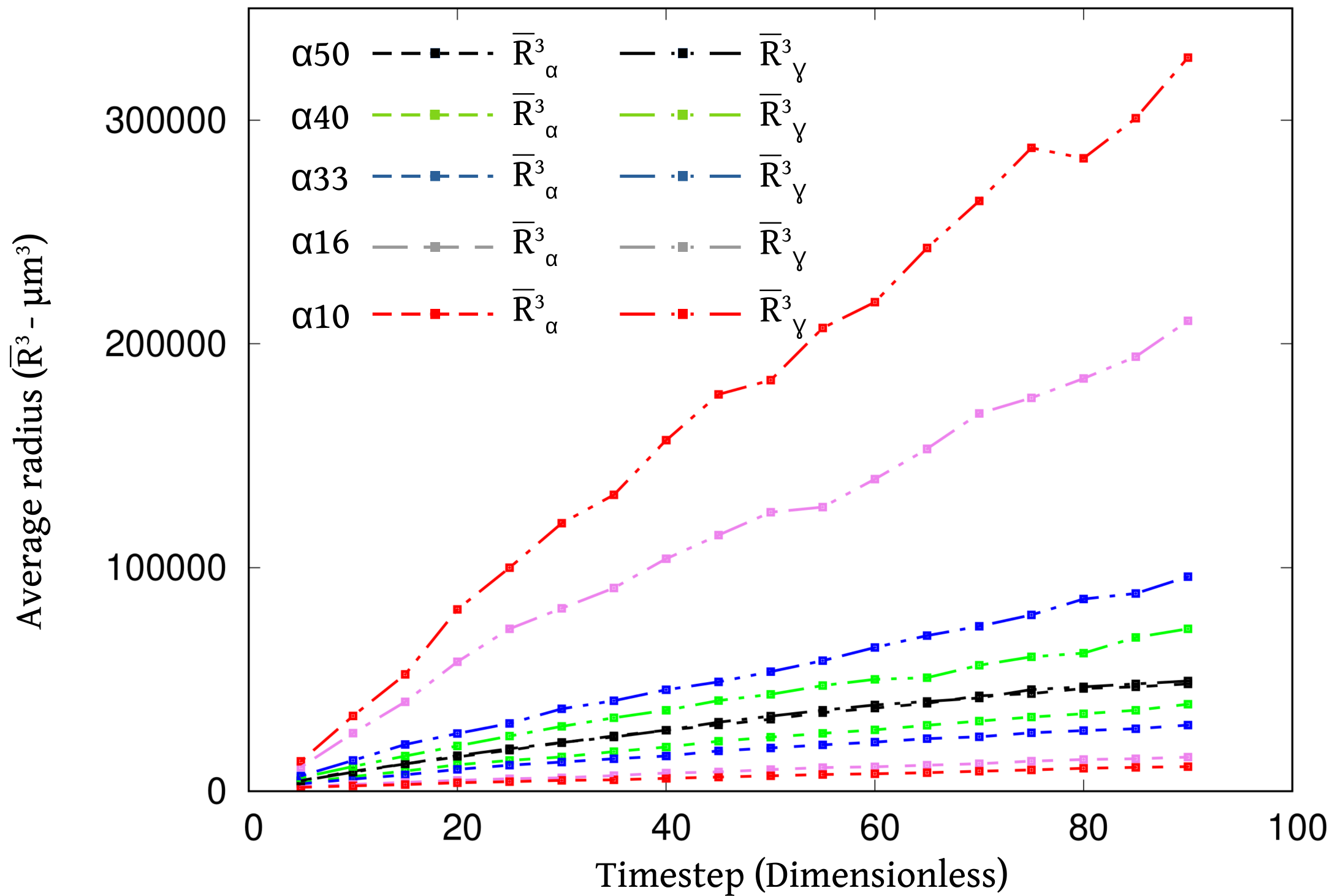}
    \end{tabular}
    \caption{  Progressive increase in the average radius of phase-$\alpha$ and -$\gamma$ grains, $\bar{R}_{\alpha}^3$ and $\bar{R}_{\gamma}^3$,  with time for various duplex systems with characteristic phase-fraction.
    \label{fig:dupR_sep}}
\end{figure}

\subsubsection{Verifying the statistical claim}

In order to substantiate the understanding rendered by the analyses based on coefficient of determination, the temporal change in the average radius of the phase-associated grains ($\bar{R}_{\alpha}$ or $\bar{R}_{\gamma}$) and entire microstructure ($\bar{R}$) are studied in a conventional manner.
In  Fig.~\ref{fig:dupR_sep}, the progressive increase in the average radius of  major- and minor-phase grains with time, in duplex systems with varying phase-fractions, are cumulatively presented. 
Since the evolution of  different duplex microstructures are considered together, for the ease of distinction, temporal change in $\bar{R}_{\alpha}^3$ and $\bar{R}_{\gamma}^3$ is adopted for this illustration.

Consistent with the mechanism of evolution, it is observed that the minor phase-grains in system with the minimal volume-fraction ($\alpha 10$) grows at a least rate.
However, the growth kinetics of the these grains noticeably increase as the corresponding phase gains more volume in the microstructure. 
Accordingly, in duplex systems with unequal phase-fractions, minor-phase grains of $\alpha 40$ microstructure exhibits highest growth-rate, followed by $\alpha 33$ and $\alpha 16$. 
On the other hand, the evolution kinetics of major-phase grains are minimal in  $\alpha 40$ system, and significantly increase in $\alpha 33$ and $\alpha 16$ as the volume fraction of the phase-$\alpha$ reduces. 
Moreover, the maximum growth-rate in $\gamma-$phase grains is observed in microstructure with minimal volume of minor phase, $\alpha 10$.
Owing to the influence of volume fraction, which governs the kinetics through the diffusion paths, the disparity in the temporal change in average radius of the major- and minor-phase grains becomes more evident as the inequality in phase-fraction increases. 
In other words, as shown in Fig.~\ref{fig:dupR_sep}, the progressive change in $\bar{R}_{\alpha}$ and $\bar{R}_{\gamma}$ with time is notably far-apart in $\alpha 10$ system when compared to the rest. 
Nevertheless, this separation gets reduced with the increase in the volume-fraction of minor phase-$\alpha$.

\begin{figure}
    \centering
      \begin{tabular}{@{}c@{}}
      \includegraphics[width=0.7\textwidth]{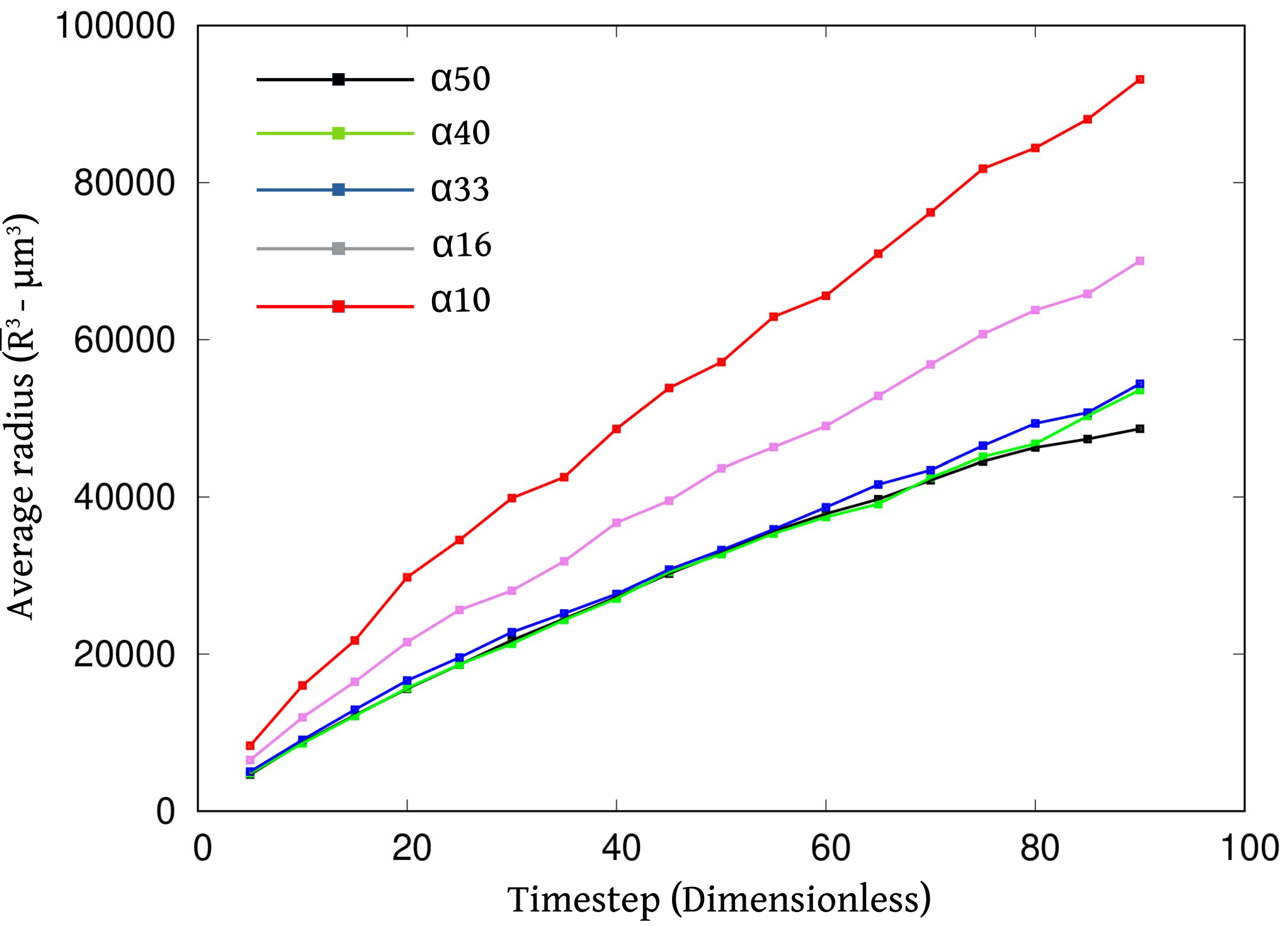}
    \end{tabular}
    \caption{ Temporal change in the average radius of entire duplex systems, $\bar{R}^3$, with varying  phase-fractions during grain growth.
    \label{fig:dupR_conso}}
\end{figure}

While Fig.~\ref{fig:dupR_sep} can be appropriately discussed to affirm the consistency of the present approach, equipped with the understanding on how evolution of individual phase-grains relate to the growth rate of entire microstructure, the general trend in the overall growth kinetics can be predicted from it. 
Analyses based on coefficient of determination, in Fig.~\ref{fig:rdq_duplex}, unravels that the growth-rate of duplex microstructure is predominantly influenced by evolution kinetics of major-phase grains. 
Moreover, the effect of the minor-phase grains decreases as their corresponding volume-fraction reduces.
Adopting these insights, and given that in Fig.~\ref{fig:dupR_sep} $\gamma$-grains of $\alpha 10$ exhibit maximum growth rate, it can be predicted that the overall growth kinetics of the corresponding microstructure will be noticeably greater than the other duplex systems considered in this study. 
Furthermore, it can also be stated that, since the volume fraction of minor-phase continues to be significantly lower in $\alpha 16$ and  $\alpha 33$, the overall growth will be dominated by the $\gamma$-phase grains, and their kinetics will correspondingly follow the $\alpha 10$ microstructure. 
Finally, considering that the volume of phase-$\alpha$ is close to major-phase in $\alpha 40$,   based on Fig.~\ref{fig:rdq_duplex}, it can be suggested that this duplex microstructure will exhibit the least rate of evolution.

In order to verify the accuracy of the above predictions, emerging from the understanding of coefficient of determination, the overall growth-rate exhibited by different duplex microstructures is cumulatively presented in Fig.~\ref{fig:dupR_conso}.
It complete adherence to the prediction, it is observed that, in duplex microstructures with unequal phase-fraction, maximum and minimum growth-rate respectively pertains to $\alpha 10$ and $\alpha 40$ microstructure. 
Additionally, the kinetics of evolution exhibited by $\alpha 16$ and  $\alpha 33$ lie in between the maximum and minimum, with the former noticeably greater than the later. 
Ultimately, Fig.~\ref{fig:dupR_conso} affirms that, in duplex systems characterised by unequal volume-fraction of constituent phases, the overall grain-growth kinetics is primarily governed by the evolution-rate of the major-phase grains.
This influence of the major-phase grains gets increasingly dominant with increase in its volume-fraction. 

\subsubsection{Phase fraction and growth rate}

\begin{figure}
    \centering
      \begin{tabular}{@{}c@{}}
      \includegraphics[width=0.7\textwidth]{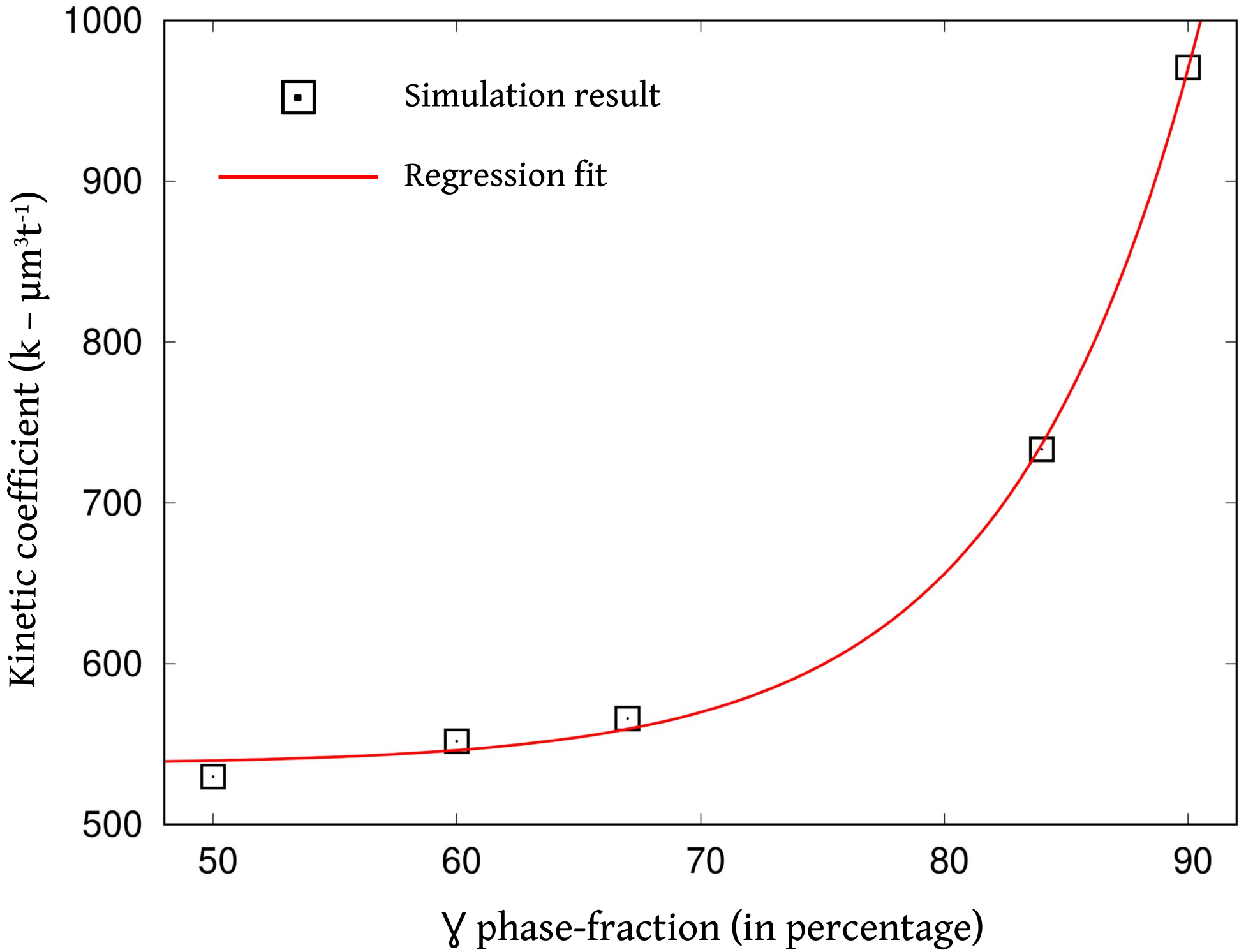}
    \end{tabular}
    \caption{  Change in the kinetics coefficients that governs grain growth with the variation in the phase-fractions of evolving duplex microstructures. 
    \label{fig:duplexK}}
\end{figure}

Having realised that, for a given duplex microstructure, the overall grain-growth rate ($d\bar{R}/dt$) is predominantly dictated by the kinetics adhere-to by the major-phase grains ($d\bar{R}_{\gamma}/dt$), attempts are made to relate the varying phase-fraction to the observed growth-rate across different systems. 
Introducing chemical-inhomogeneity amongst the grains, in a polycrystalline system, shifts the governing mechanism, irrespective of the volume-fraction of the phases. 
Kinetics rendered by this mechanism in multiphase microstructure complies with the power-law characterised by the exponent $n=3$.
Considering that the exponent remains unaltered reflecting the governing mechanism, despite the varying phase-fraction in multiphase systems, the disparity in the rate of grain growth can only be understood from the kinetic coefficient, $k$, that relates the average radius to time. 

By studying the evolution of different duplex systems considered in this study, the corresponding kinetic coefficients are ascertained. 
Kinetic coefficients associated with duplex microstructures with varying phase-fractions are illustrated in Fig.~\ref{fig:duplexK}. 
Since the growth rate of the duplex microstructure are principally dictated by the evolution of major-phase grains, the corresponding volume-fraction is considered for this representation. 
Fig.~\ref{fig:duplexK} indicates that with increase in the volume of a major phase in duplex system, grain growth in the system occurs at an higher rate. 
In order definitively understand the influence of the $\gamma$-fraction on the kinetic coefficient exhibited by the corresponding duplex microstructure, non-linear regression technique is adopted and a relation expressed as 
\begin{align}\label{Dpf_k}
k_{\text{dup}} = A_{\text{dup}} + B_{\text{dup}}\exp(C_{\text{dup}}\dot V_{\gamma}),
\end{align}
is realised, where $V_{\gamma}$ is the volume fraction of the major phase-$\gamma$.
The constants $A_{\text{dup}}$, $B_{\text{dup}}$ and $C_{\text{dup}}$, for the present consideration, respectively assume the value of $537.3$, $3.8\times 10^{-3}$, and $0.13$. 
In the above relation, it is vital to note that $k_{\text{dup}}$ indicates the kinetic coefficient adhered-to by the entire duplex system during grain growth, not the evolving major- or minor-phases. 

\subsection{Triplex microstructures}

Triplex systems are characterised by the association of  individual grains, in the polycrystalline setup, to one of the three constituent phases. 
Corresponding microstructure, in this study, comprises of phases $\alpha$, $\beta$ and $\gamma$, with $\gamma$ largely acting as the matrix or major-phase. 
In the existing works, unlike duplex systems, very few three-phase microstructures with varying phase-fractions are analysed~\cite{ravash2017three2}.
This limited consideration of triplex microstructure can largely be attributed to the computational burden associated with it. 
Moreover, conventionally, the grain-growth kinetics of the triplex microstructure are discussed by focusing on the evolution of the individual phase-grains without sufficiently relating it the overall system.
In the present study, on the other hand, grain growth in fourteen different three-phase microstructures, with varying phase-fractions, are examined to elucidate with statistical certainty how the evolution of the individual phase-grains effects the growth kinetics of entire triplex microstructure. 

\subsubsection{Grain growth kinetics}

\begin{figure}
    \centering
      \begin{tabular}{@{}c@{}}
      \includegraphics[width=0.75\textwidth]{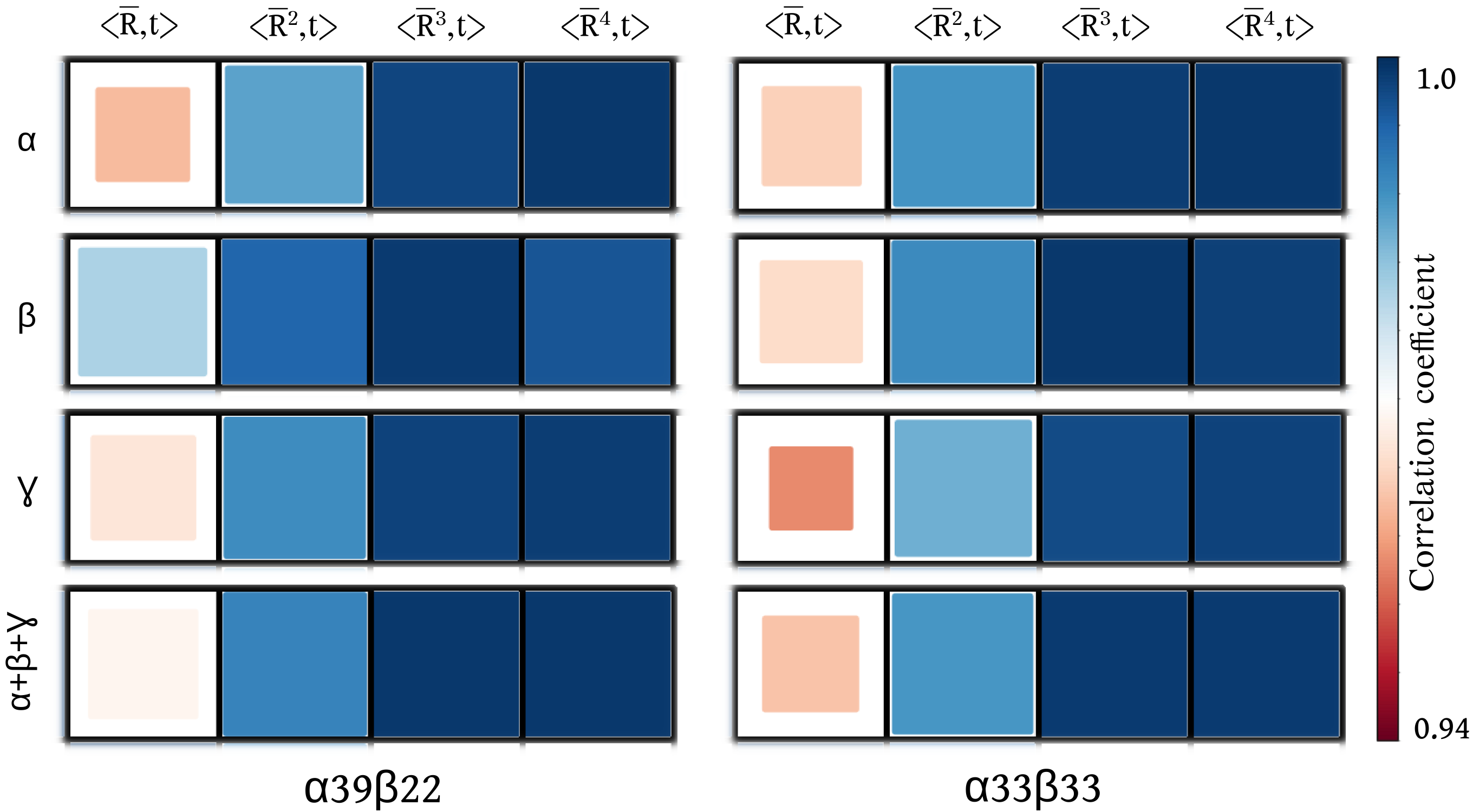}
    \end{tabular}
    \caption{  Correlation coefficient characterising the relation between time and average radius of individual phase-grains and entire microstructure, raised to different power ($n=1,2,3$ and $4$), for two triplex systems, $\alpha39 \beta22$ and $\alpha33\beta33$. 
    \label{fig:triplex_power}}
\end{figure}

Despite the difference in the number of phases, grain growth in both duplex and triplex systems are fundamentally governed by the same mechanism. 
Therefore, the grain growth in three-phase systems, which is dictated by the diffusion of chemical components, adheres to the power law with the exponent, $n=3$. 
In order to ensure that the grain growth in triplex systems is accurately modelled by the present approach, a validation technique adopted for duplex microstructures, is extended. 
Correspondingly, the temporally-varying average radius of the individual phase-grains, $\bar{R}_{\alpha}$, $\bar{R}_{\beta}$ and  $\bar{R}_{\gamma}$, and overall microstructure, $\bar{R}$, are raised to different powers ($n=\{1,2,3,4\}$) and related to time, $t$.
Correlation coefficient characterising the different relations are ascertained for two triplex microstructures, $\alpha33\beta33$ and $\alpha39\beta22$,  and are graphically illustrated in Fig.~\ref{fig:triplex_power}. 

It is evident in Fig.~\ref{fig:triplex_power} that correlation coefficient relating the average radius to the time is highest when $n=3$ for both individual phases, and overall microstructure.
The maximum correlation exhibited by the cube of the different average radii, $\bar{R}_{\alpha}$, $\bar{R}_{\beta}$,  $\bar{R}_{\gamma}$ and $\bar{R}$, with time implies that the growth of the individual phase-grains, and the entire microstructure, are predominantly governed by the long-range diffusion of the chemical components. 

\subsubsection{Ascertaining governing factor}

Considering that grain growth in both duplex and triplex system are predominantly dictated by the diffusion of the chemical components, phase-fraction renders identical influence on the evolution of individual phases. 
In other words, when certain phase(s) assumes minor volume-fraction in the three-phase microstructure, owing to relative increase in the length, and complexity, of the diffusion path, the growth of the corresponding grains are stunted. 
On the other hand, the evolution kinetics is enhanced when the volume of the phase(s) is dominant in the multiphase systems. 
Apart from these generalised understanding, existing report rarely offer any further insights on the grain-growth kinetics in triplex microstructures.
Particularly, similar to duplex system, sufficient consideration has not been rendered to relate the growth kinetics of the individual phases to the evolution of the entire triplex system. 
To that end, in this analysis, the impact of the growth rate of  individual phase-grains on that entire three-phase microstructure is examined by ascertaining the corresponding coefficient of determination. 

\begin{figure}
    \centering
      \begin{tabular}{@{}c@{}}
      \includegraphics[width=1.0\textwidth]{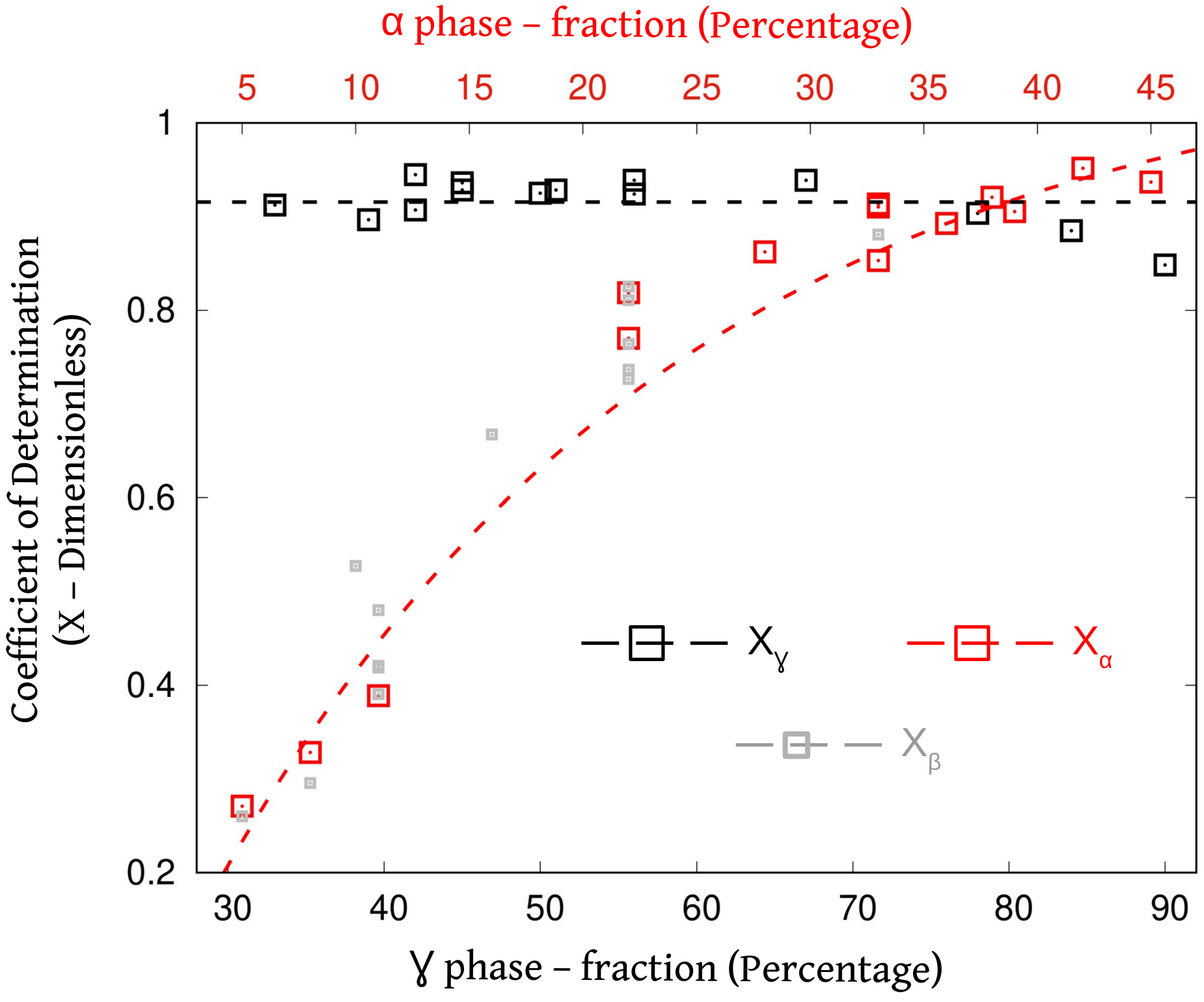}
    \end{tabular}
    \caption{  Change in coefficient of determination, typifying the influence of growth rate of individual phase-grains on the evolution kinetics of entire microstructure, with variation in the volume fraction of constituent phases.
    \label{fig:triALL_Rsq}}
\end{figure}

The instantaneous growth rate for constituent phase-grains, $d\bar{R}_{\alpha}/dt$, $d\bar{R}_{\beta}/dt$ and $d\bar{R}_{\gamma}/dt$, along with the entire triplex microstructure, $d\bar{R}/dt$, is determined by monitoring temporal change in the respective parameter. 
These instantaneous growth kinetics of the individual phase-grains are related to that of the entire microstructure, and the corresponding coefficient of determination is estimated through the Eqn.~\eqref{coeff_deter}.
For each system, three distinct coefficients of determination, $\chi_{alpha}$,  $\chi_{beta}$ and $\chi_{gamma}$, are estimated, reflecting the characteristic feature of the triplex microstructure. 
These coefficients of determinations are related to the phase-fractions of the microstructure, and illustrated in  Fig.~\ref{fig:triALL_Rsq}. 

In the triplex systems considered in the present study, the volume-fraction of phase-$\gamma$ reaches as low as 33\% despite being the major phase. 
Such volume-fraction of phase-$\gamma$ is noticed in triplex microstructure characterised by equifraction of phases. 
Furthermore, in some system like $\alpha 45 \beta 10$, the phase-$\alpha$ assumes a volume-fraction of $45\%$, in spite of being one of the minor phases.
This, and similar, understanding of phase-fraction is vital to investigate the coefficients of determination presented in Fig.~\ref{fig:triALL_Rsq}. 

Fig.~\ref{fig:triALL_Rsq} unravels that, even though the volume-fraction of the major-phase varies noticeably across different triplex systems, the corresponding coefficient of determination, $\chi_{gamma}$, continues to remain the high. 
Given that phase-$\gamma$ stays as a major-phase, despite the change in phase-fractions, the high values of $\chi_{gamma}$ can be attributed to the dominant volume of the respective grains.
In other words, analogous to the duplex microstructure, the evolution kinetics of major-phase grains offer relatively greater influence on the overall growth-rate exhibited by the entire triplex microstructure. 
Furthermore, Fig.~\ref{fig:triALL_Rsq} suggests that  the coefficient of determination of the minor-phases, $\alpha$ and $\beta$, noticeably increases as their corresponding volume-fraction raises. 
Particularly, as the volume of phase-$\alpha$ gets as dominant as $\gamma$, in a triplex system, identical coefficient of determination is rendered, $\chi_{alpha} = \chi_{gamma}$.
On the other hand, when the volume of the phases are minimal, the respective coefficient of determination assumes least value. 
Ultimately, it is evident from Fig.~\ref{fig:triALL_Rsq} that the influence of the individual phase-grains on the overall growth kinetics depends largely on the corresponding volume-fraction. 
In a triplex system with unequal volume-fraction of phases, the growth rate of the entire microstructure, $d\bar{R}/dt$, is predominantly governed by the evolution of the major-phase grains, $d\bar{R}_{\gamma}/dt$.
When volume of two phases are dominant in a three-phase system, the growth rate of both these phase-grains offer identical influence on evolution of the microstructure. 
The contribution of a given phase-grains to the overall evolution kinetics, $d\bar{R}/dt$, becomes least, when its volume-fraction are minimal. 
Based on the understanding rendered by Fig.~\ref{fig:triALL_Rsq},  as demonstrated for duplex systems (Fig.~\ref{fig:dupR_conso}), the growth-kinetics of a triplex microstructure, in relation to others with varying phase-fractions, can be predicted from the temporal change in the average radius of the corresponding phase-associated grains. 

\subsubsection{Interdependency in the evolving phases}

In duplex systems, since the grains of the polycrystalline microstructure are associated with either of the two  constituent phases, the evolution of a particular phase-grains, and its kinetics, are inherently bound to the other. 
However, the same interdependency cannot be expected in triplex systems, wherein the grains can be associated with one of the three possible phases. 
Moreover, in three-phase microstructures, level of influence offered by one evolving phase-grains on the rest of the phase-associated grains has not been conscientiously addressed yet.
Therefore, by examining the temporal change in the average radius of a particular phase-grains in relation to the others, the interdependency exhibited between the phases, in triplex microstructures, during grain growth is elucidated. 

\begin{figure}
    \centering
      \begin{tabular}{@{}c@{}}
      \includegraphics[width=1.0\textwidth]{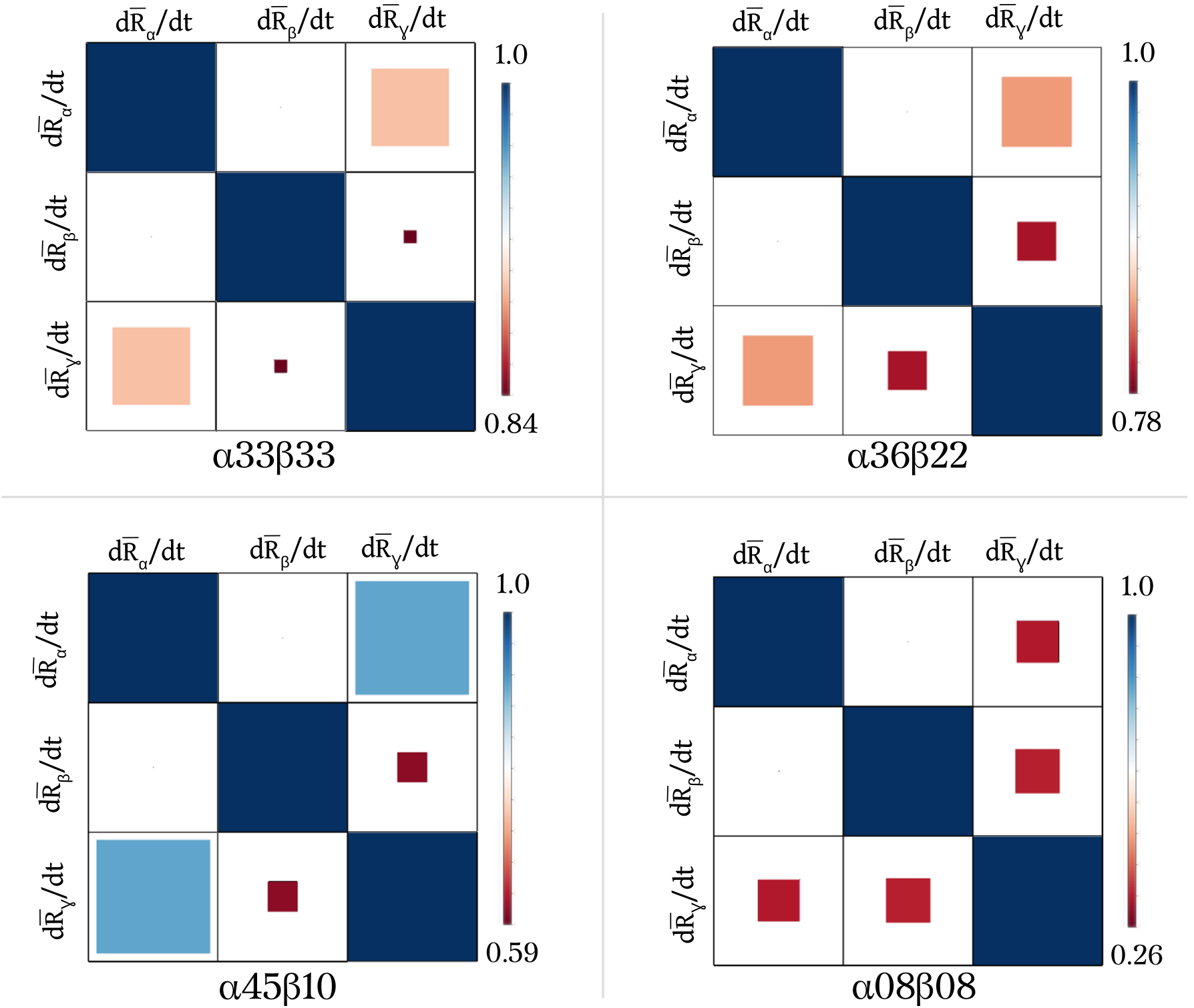}
    \end{tabular}
    \caption{  Correlation coefficient explicating the interdependency between the growth rate of different phase-grains, $d\bar{R}_{\alpha}/dt$, $d\bar{R}_{\beta}/dt$ and $d\bar{R}_{\gamma}/dt$, during grain growth of four different triplex system with varying phase-fractions. 
    \label{fig:rate_corr}}
\end{figure}

Instead investigating all the fourteen triplex microstructure to explicate the effect of one evolving phases on the other, the systems are categorised based on phase-fraction, and one microstructure from each category is analysed. 
Since the coefficient of determination, which quantifies the effect of  an evolving-phase grains on the growth of the overall triplex microstructure, depends on volume fraction, the phase-fraction based grouping is deemed reasonable. 
Apart from the equifraction system, $\alpha 33 \beta 33$, wherein all the constituent phases largely occupy similar volume, the remaining systems can be categorised as \lq equi-major \rq \thinspace, \lq equi-minor \rq \thinspace and \lq non-equifraction \rq \thinspace triplex microstructures. 
While, in equi-major system, volume of one of the minor phase is equal to that of the major-phase  ($V_\alpha= V_\gamma$), the volume fraction of minor phases are identical in equi-minor microstructures ($V_\alpha= V_\beta$).
Moreover, the non-equifraction system stand in direct contrast to the equifraction microstructure, and characterised by totally unequal volume-fraction of the constituent phases ($V_\alpha\neq V_\beta \neq V_\gamma$). 

In order understand the degree of interdependency between the evolving phases, in addition to equifraction microstructure, grain growth exhibited systems  $\alpha 36 \beta 22$, $\alpha 45 \beta 10$ and $\alpha08\beta08$ pertaining to non-equifraction, equi-major and equi-minor, respectively, are analysed.
Given that the primary focus of  the present investigation is \textit{not} to quantify the effect of an evolving-phase grains on the rest, but rather to qualitatively realise the degree of interaction between two phases, during grain growth in a triplex systems, coefficient of determination is not estimated. 
However, alternatively, the growth rate of different phase-associated grains are estimated, $d\bar{R}_{\alpha}/dt$, $d\bar{R}_{\beta}/dt$ and $d\bar{R}_{\gamma}/dt$, and are related to  (or plotted against) each other.
Correlation coefficient characterising the relation between the growth rate of two phase-grains are realised for pre-determined triplex microstructures, and graphically illustrated in Fig.~\ref{fig:rate_corr}. 

Before elucidating level of interaction between two evolving phase-grain in a given triplex microstructure, based on Fig.~\ref{fig:rate_corr}, it is exceedingly critical to realise the variation in the range of correlation coefficient across the different systems. 
Particularly, as opposed to the maximum value, which remains constant at unity, the least value of the correlation coefficient changes with phase-fraction. 
Correspondingly, Fig.~\ref{fig:rate_corr} unravels that the lowest correlation-coefficient in the equifraction system is maximum ($0.84$) when compared to the rest of the triplex microstructures, and it is respectively followed by non-equifraction ($\alpha 36 \beta 22$) and equi-major ($\alpha 45 \beta 10$) microstructures, with the absolute minimal exhibited by the equi-minor system ($\alpha08\beta08$). 
This significant disparity in the lowest value of correlation coefficient emphasis the importance of considering the context, $i.e$ the correlation-coefficient range, while interpreting the interaction between two-phases in a given microstructure during grain growth.
In other words, apparently least interdependency between the two-phase grains in a equifraction system would translate to a strong interaction in the context of equi-minor triplex microstructure. 

The graphical representation of correlation coefficient, in Fig.~\ref{fig:rate_corr}, that indicates level of influence the growth rate of one evolving phase-grain has on the other,  during grain growth in triplex microstructures, unravels few similarities and dissimilarities across the systems with varying phase-fractions. 
It is evident in this illustration that, irrespective of the nature of the three-phase system, the correlation coefficient relating the growth rate of $\alpha$- and $\beta$-phase grains are minimal, within a given microstructure. 
In other words, during grain growth in a triplex system, the evolution kinetics of one minor-phase grains imposes the least effect on growth rate of other low-volume phase-grains.
Despite being equifraction, largely owing the manner in which the triplex microstructure is initialised, such effect is also observed in $\alpha 33 \beta 33$ microstructure.
However, it is vital to note that the least correlation in equifraction system is tantamount to noticeable interaction in relation to other triplex microstructures. 
Therefore, in  $\alpha 33 \beta 33$ system, the evolution kinetics of one-phase grains is generally interlinked with grains of the other phases, but this interaction is least between phase-$\alpha$ and -$\beta$.

Within a triplex system, during grain growth, Fig.~\ref{fig:rate_corr} suggests that minimal interdependency following two minor-phases is observed between the minor-phase and the matrix grains, irrespective of the phase-fractions. 
The only exception is the equi-minor system wherein the volume-fraction of the minor-phases are identical. 
Furthermore, in all triplex systems, the growth rate of minor-phase grains with relatively greater volume-fraction when compared to the other ($V_{\alpha} > V_{\beta}$ ) is strongly coupled with the evolution of the major-phase grains. 
Ultimately, the study of interdependency between the rate of the evolving phase-grains during grain growth in triplex systems, using correlation coefficients, unravel that a general trend is observed in three-phase microstructures irrespective of the phase-fractions. 
If we distinguish the constituent phases as minor- , inter- and major-phase depending on their corresponding volume fraction, which in the present study is $\alpha$,$\beta$ and $\gamma$ respectively, then the least interaction during grain growth is exhibited by the minor- and inter-phase grains.
While the growth rate of major-phase grains are considerably interlocked with the inter grains, the effect of the minor phase on the matrix is comparatively lower.
In other words, in a given triplex microstructure, the level of influence offered by the evolution rate of one phase-grains on the other, during grain growth, is primarily dictated by the their corresponding volume fractions. 
When the volume fraction of two phases are minimal, their degree of interaction is also minimal, whereas a considerable dependency is noticed when the volume of the two phases are dominant in view of the third. 

This analysis on the interdependency of the kinetics of evolving phases in triplex system, during grain growth, unravels that, for expressing the overall growth rate of a three-phase microstructure, the two ideal variables, with least \textit{multicollinearity}, are the evolution rate of the grains of minor phases. 

\begin{figure}
    \centering
      \begin{tabular}{@{}c@{}}
      \includegraphics[width=0.9\textwidth]{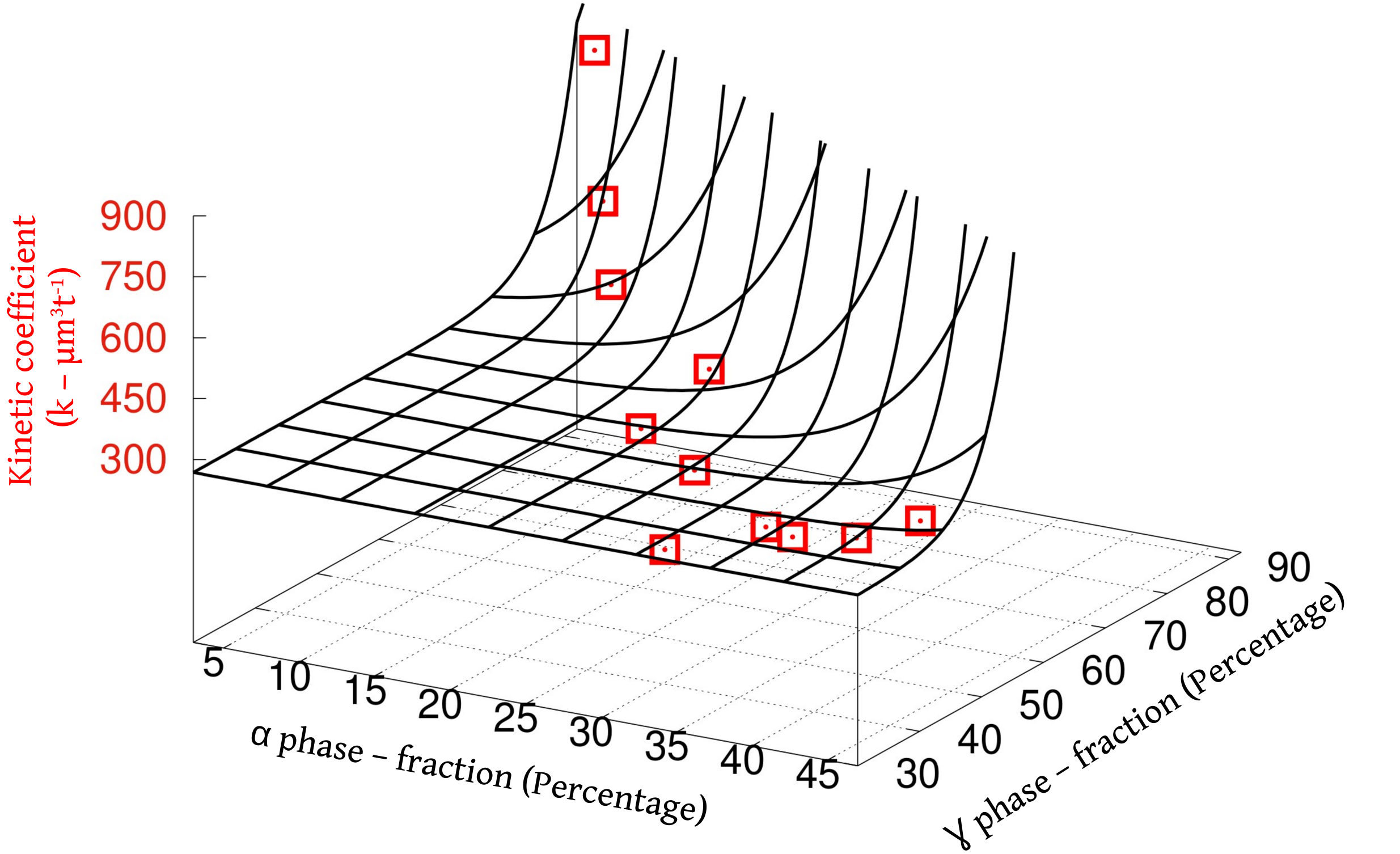}
    \end{tabular}
    \caption{  Effect of phase-fraction on the kinetic coefficient governing the rate of grain growth in triplex systems.
    \label{fig:triALL_fitting}}
\end{figure}

\subsubsection{Phase fractions and growth rate}

A principal insight rendered by the present study is that, in grain growth, the evolution rate of a multiphase system is primarily governed by the volume-fraction of its constituent phases through their respective growth kinetics. 
Therefore, by relating the phase-fraction of  fourteen different triplex microstructures, considered in this work, to its corresponding growth kinetics, an attempt to extract a generalised expression. 
Unlike in duplex system, since triplex microstructures are characterised by three constituent phases,  the corresponding evolution rate is dictated by two independent variables, $i.e$ volume fractions. 
By directly relating the volume fraction, instead of the evolution rate of individual phase-grains, to the growth kinetics exhibited by the triplex systems, the question of multicollinearity is obviated.

Based on the proportionality exhibited by the phase-fraction and growth kinetics in duplex microstructure, multiple non-linear regression is employed to relate the volume-fraction of constituent phases to the corresponding grain growth-rate exhibited by the entire three-phase microstructures. 
Strictly owing to the different phase-fractions considered in the present analysis, volume fraction phase-$\alpha$ and -$\gamma$ are considered as the independent variables. 
Expression rendered by multiple non-linear regression that relates phase-fractions to the kinetic coefficient governing the grain-growth rate of  entire three-phase microstructure reads
\begin{align}\label{Tpf_k}
 k_{\text{tri}}=A_{\text{tri}} + B_{\text{tri}} \exp(C_{\text{tri}}^{\alpha}V_{\alpha}+C_{\text{tri}}^{\gamma}V_{\gamma}),
\end{align}
where $V_{\alpha}$ and $V_{\gamma}$ correspond to the volume fraction of phase-$\alpha$ and -$\gamma$.
Moreover, $A_{\text{tri}}$, $B_{\text{tri}}$, $C_{\text{tri}}^{\alpha}$ and $C_{\text{tri}}^{\gamma}$ are constants whose respective values are $268.2$, $2.2 \times 10^{-7}$, $0.19$ and $0.22$. 
In Fig.~\ref{fig:triALL_fitting} kinetic coefficient adhered to different triplex microstructures during grain growth is plotted against the corresponding volume fraction of phase-$\alpha$ and -$\gamma$.
The curve reflecting Eqn.~\eqref{Tpf_k} is included in this illustration.
Noticeably good agreement between the numerical results and the curve indicates that Eqn.~\eqref{Tpf_k} convincingly relates the kinetic coefficients characterising the grain growth in isotropic triplex systems to the phase-fractions.  

\section{Conclusion}

Grain growth in polycrystalline systems can be desirous under certain conditions, while unwelcomed in others. 
For instance, grain growth is induced during processing technique to establish required average grain-size, while noticeable measures are generally taken to avoid it during a given application. 
The subjective role of grain growth extends beyond homogeneous polycrystalline system to multiphase microstructures as well. 
Therefore, it becomes vital to understand the grain-growth kinetics exhibited by highly-applicable multiphase polycrystalline microstructures associated with duplex and triplex systems. 
Particularly, generalised insights that aide in comprehending the growth rate of multiphase microstructures with varying phase-fraction are exceedingly critical, as they can be adopted for wide-range of systems and application. 
To that end, in this study, the grain-growth kinetics of duplex and triplex systems are studied by employing approachable statistical techniques. 

Conventionally, the grain-growth kinetics associated with multiphase systems are discussed by considering the evolution rate of individual phase-grains and entire-microstructure separately. 
Such treatments rarely offer much insights on how the growth kinetics of individual phases relate to the overall evolution rate exhibited by the entire system. 
Therefore, in the present work, statistical tools are employed to realise the effect of growth rate of a given phase-associated grains on the overall kinetics of evolving microstructure. 
Correspondingly, it is unraveled that, during grain growth in multiphase systems, the influence of the growth kinetics of a phase-grains on the evolution of entire system depends on its volume fraction. 
In other words, in systems with varying volume fraction of constituent phases, the evolution kinetics of the major-phase grains predominantly govern the overall growth rate exhibited the microstructure. 
Even though, when focusing on a specific system, it might appear that the temporal change in the average radius of the entire multiphase system lies close the evolution of the corresponding radius of the minor-phase grains, when viewed in relation to microstructures of varying phase-fractions, the dominance of growth kinetics of the major-phase grains gets increasingly evident. 
This holistic understanding that encompasses multiphase systems with varying phase-fractions, though arrived from  statistical treatments, in the present work, it is vindicated through conventional representation and corresponding discussions.
Moreover, in addition to effecting the role of individual phase-grains on the evolution kinetics of entire multiphase system, phase fraction also dictates the interdependency between growth rates of different phase-grains. 
During grain growth, evolution kinetics of grains of two phases are largely independent when their corresponding volume-fractions in the multiphase system are minimal.
On the other hand, noticeable interaction is observed in the grain-growth rate of two phases that dominant the multiphase microstructure. 

Understanding rendered by the present investigation can be exploited for various purposes, however, one critical utilisation would be to alter the grain-growth rate of a given multiphase system, with a definite phase-fraction, by appropriately, and exclusively, varying the evolution kinetics of the major-phase grains. 
To that end, in the upcoming works, attempts will be made to substantiate the approach of modifying the grain-growth rate in multiphase systems by employing the dominant influence of the major-phase grains.

\section*{Appendices}

\subsection*{Appendix 1: Interdependency in duplex systems}\label{sec:app1}

\begin{figure}
    \centering
      \begin{tabular}{@{}c@{}}
      \includegraphics[width=1.0\textwidth]{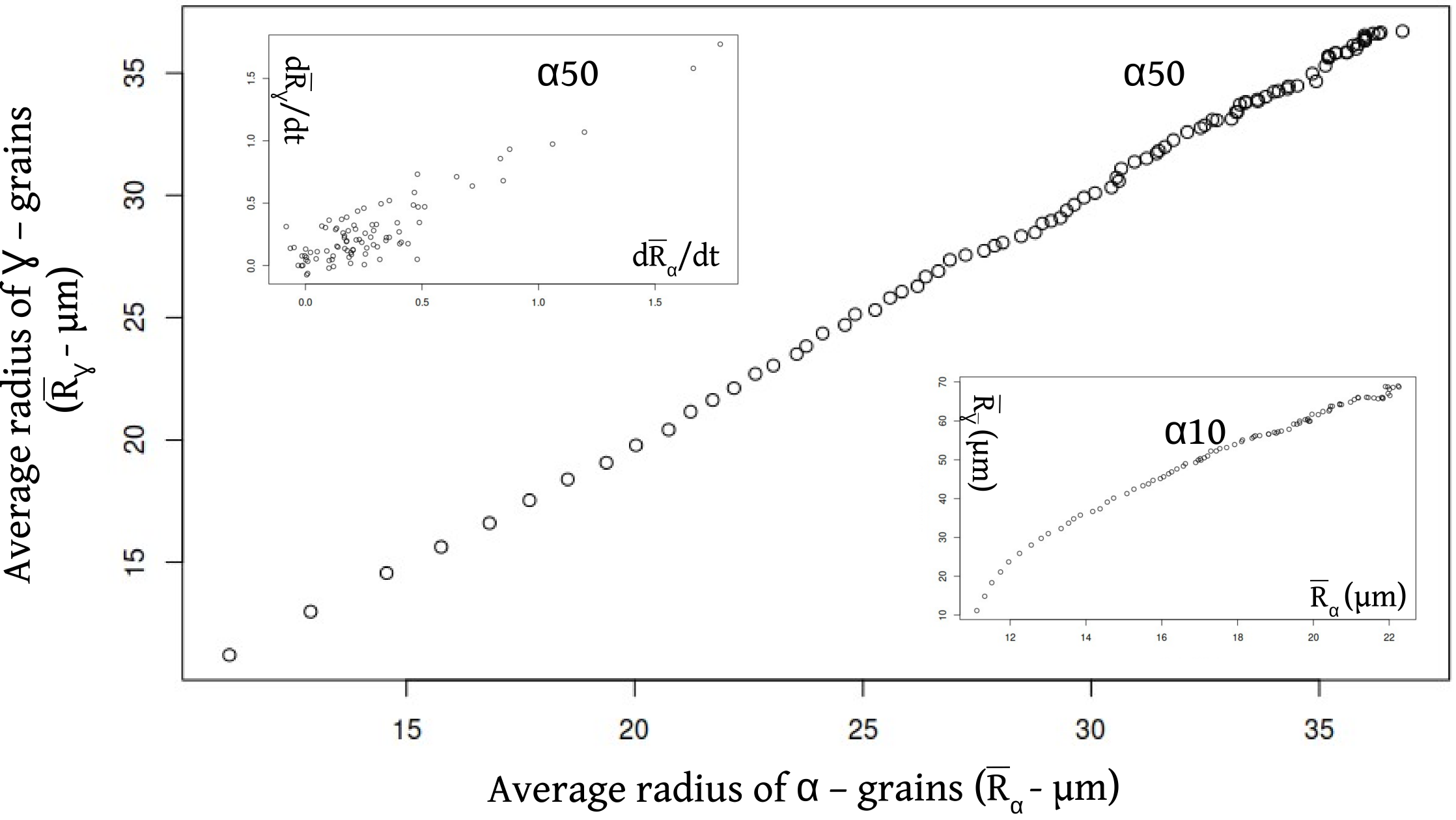}
    \end{tabular}
    \caption{  The average radius of phase-$\alpha$ grains, at a given instance, is plotted against the corresponding radius of matrix-phase grains for duplex microstructure with equal volume-fraction of phases. In the subplots, for the same system, the growth rate of individual phase-grains are related, and instantaneous average radius of phase-$\alpha$ and -$\gamma$ grains are plotted for duplex microstructure with $10\%$ alpha.
    \label{fig:app1}}
\end{figure}

In a polycrystalline system, irrespective of its nature, the continuum is established by the multiple grains present it. 
During the grain growth, despite the continual disappearance of the grains, the continuum is sustained by the growth the surviving grains. 
This characteristic feature of the grain growth introduces interdependency between the evolving grains. 
Correspondingly, in duplex systems, wherein the grains are associated with one of the two-constituent phases, the evolution of phase-$\alpha$ grains are inherently linked the grains of phase-$\gamma$. 
Even though the interaction between the phase-associated grains, during grain growth of a duplex system, can be theoretically conceived, to explicate it with a statistical certainty, the average radius of $\alpha$-grains, at a given time $t$, is plotted against corresponding of radius of $\gamma$ grains in Fig.~\ref{fig:app1} for two-phase microstructure with equal volume-fraction of phases, $\alpha 50$.
The trend in this illustration indicates a inherent interlocking between the evolution of the phase-$\alpha$ and -$\gamma$ grains during the grain growth of equifraction duplex-microstructure.
In addition to the $\alpha 50$ microstructure, the average radius of constituent phase-grains, at a given instance, are ascertained for duplex system with $10\%$ minor phase-$\alpha$.
Similar to equifraction system, these instantaneous average radii of phase-$\alpha$ and -$\gamma$ grains are plotted against each other, and illustrated in Fig.~\ref{fig:app1} as a subplot.
Despite the change in the phase-fraction, in $\alpha 10$microstructure as well, a definite interaction between the radius of major- and minor-phase grains is evident. 

Besides the average radius of the phase-$\alpha$ and -$\gamma$ at a given instance, using the same approach, the relation between the kinetics of grain growth associated with these phases can be explicated. 
Correspondingly, the evolution kinetics of the $\alpha$-grains are related to that of the $\gamma$ ones, for equifraction duplex microstructure, and are included as a subplot in Fig.~\ref{fig:app1}. 
This illustration unravels that even though there exists a perceivable interdependency between the instantaneous kinetics of phase-$\alpha$ and -$\gamma$ grains, it is not as straightforward as the average radius
 
\subsection*{Appendix 2: Effect of individual phase-grains kinetics on growth-rate of entire system}\label{sec:app1}
 
 \begin{figure}
    \centering
      \begin{tabular}{@{}c@{}}
      \includegraphics[width=1.0\textwidth]{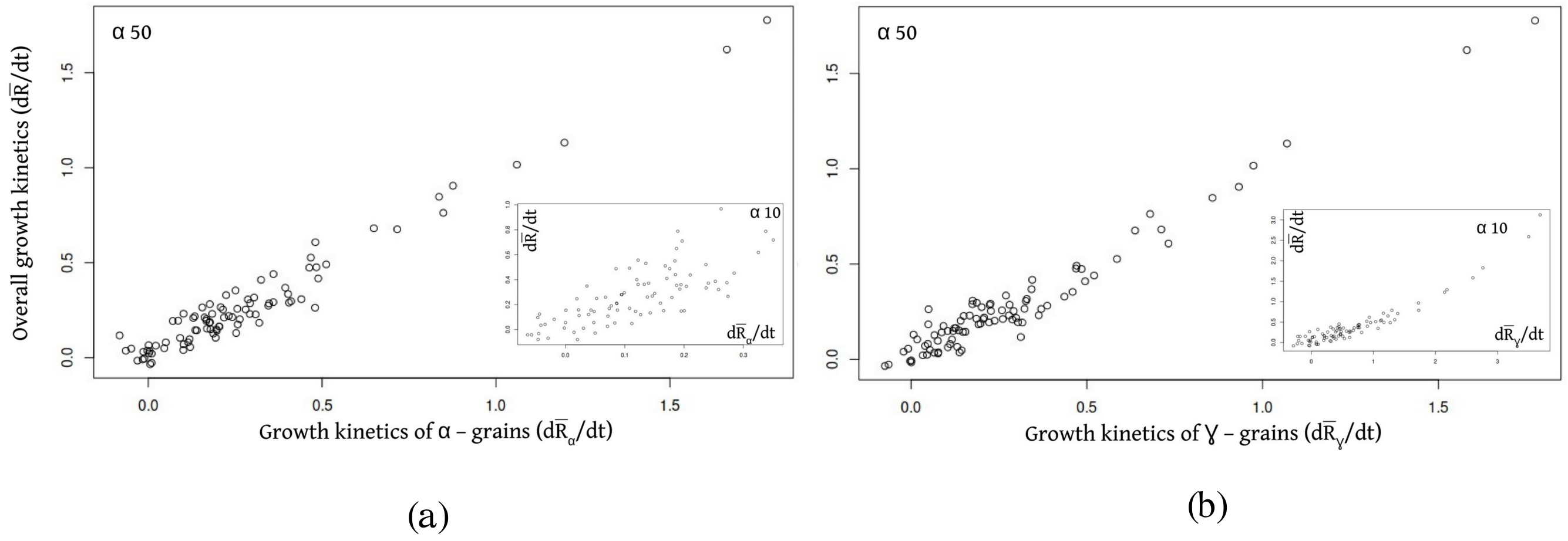}
    \end{tabular}
    \caption{  The instances grain-growth rate exhibited by an individual phase-grains are plotted with respect to that of the entire duplex microstructure with equal volume fraction of phases. In (a) the kinetics of $\alpha$-grains are related to the grain-growth rate of the entire microstructure, while $\gamma$-grains evolution kinetics are considered (b). The corresponding outcomes for $\alpha 10$ microstructure with $10\%$ of minor phase-$\alpha$ are included as subplots. 
    \label{fig:app2conso}}
\end{figure}

One of the primary aim of the present investigation is to realise the effect of individual phases on grain-growth rate of entire two-phase microstructure. 
Particularly, the role of evolution kinetics of a given phase-grains on the overall growth rate of duplex system. 
To that end, the kinetics of evolution exhibited by phase-$\alpha$ grains, at a given instance, is related to the overall growth rate of equifraction duplex-system, and plotted in Fig.~\ref{fig:app2conso}a. 
The graphical illustration relating the evolution rate of  phase-$\alpha$ grains and duplex microstructure with $10\%$ minor phase is included as a subplot. 

Fig.~\ref{fig:app2conso}a unravels that the growth rate of phase-$\alpha$ grains imposes a definite influence on the overall kinetics exhibited by the equifraction-duplex system, $\alpha 50$, during grain growth. 
On the other hand, the subplot of the corresponding illustration, which pertains to two-phase microstructure with $10\%$ of phase-$\alpha$, indicates a relation between $d\bar{R}_{\alpha}/dt$ and $d\bar{R}/dt$, it is not as definite as one noticed in the equifraction microstructure.  
It is the degree of inter-relation between the kinetics of individual phase-grains, and overall growth-rate of a given duplex microstructure, which varies with the phase-fraction, is realised by \textit{coefficient of determination}. 
In other words, while the coefficient of determination relating the  $d\bar{R}_{\alpha}/dt$ and $d\bar{R}/dt$ will be higher for equifraction duplex microstructure, in $\alpha 10$ system it will assume a relatively low value reflecting a not so definite relation between the kinetics.

In Fig.~\ref{fig:app2conso}b the grain-growth kinetics of phase-$\gamma$ grains and entire duplex system, with equal volume-fraction of phases, is plotted against each other. 
As a subplot, the corresponding relation between $d\bar{R}_{\gamma}/dt$ and $d\bar{R}/dt$ for $\alpha 10$ duplex microstructure with $10\%$ of phase-$\alpha$ is illustrated. 
Unlike the influence of phase-$\alpha$ on overall growth-kinetics in Fig.~\ref{fig:app2conso}a, a highly definite relation is observed between the evolution rate of phase-$\gamma$ grains and entire microstructure in both equifraction and $\alpha 10$ duplex microstructure. 
Consequently, the corresponding values of coefficient of determination will largely be independent of the phase-fraction.

\bibliographystyle{elsarticle-num}
\bibliography{library.bib}

\begin{thebibliography}{10}

\bibitem{jacques2001developments}
PJ~Jacques, E~Girault, PH~Harlet, and Francis Delannay.
\newblock The developments of cold-rolled trip-assisted multiphase steels. low
  silicon trip-assisted multiphase steels.
\newblock {\em ISIJ international}, 41(9):1061--1067, 2001.

\bibitem{huang2017multiphase}
Yanli Huang, Chunlin Zhao, Xiang Lv, Hui Wang, and Jiagang Wu.
\newblock Multiphase coexistence and enhanced electrical properties in (1-xy)
  batio3-xcatio3-ybazro3 lead-free ceramics.
\newblock {\em Ceramics International}, 43(16):13516--13523, 2017.

\bibitem{kumar2020carbon}
Amit Kumar, Kamal Sharma, and Amit~Rai Dixit.
\newblock Carbon nanotube-and graphene-reinforced multiphase polymeric
  composites: review on their properties and applications.
\newblock {\em Journal of Materials Science}, 55(7):2682--2724, 2020.

\bibitem{zhang2011microstructure}
Xiaodan Zhang, Andy Godfrey, Xiaoxu Huang, Niels Hansen, and Qing Liu.
\newblock Microstructure and strengthening mechanisms in cold-drawn pearlitic
  steel wire.
\newblock {\em Acta Materialia}, 59(9):3422--3430, 2011.

\bibitem{hwang2019microstructure}
Joong-Ki Hwang.
\newblock The microstructure dependence of drawability in ferritic, pearlitic,
  and twip steels during wire drawing.
\newblock {\em Journal of Materials Science}, 54(11):8743--8759, 2019.

\bibitem{liljas2008development}
Mats Liljas, Pelle Johansson, Hui-Ping Liu, and Claes-Olof~A Olsson.
\newblock Development of a lean duplex stainless steel.
\newblock {\em steel research international}, 79(6):466--473, 2008.

\bibitem{armas2002mechanical}
AF~Armas, C~Petersen, R~Schmitt, M~Avalos, and I~Alvarez-Armas.
\newblock Mechanical and microstructural behaviour of isothermally and
  thermally fatigued ferritic/martensitic steels.
\newblock {\em Journal of nuclear materials}, 307:509--513, 2002.

\bibitem{filip2003effect}
R~Filip, K~Kubiak, W~Ziaja, and J~Sieniawski.
\newblock The effect of microstructure on the mechanical properties of
  two-phase titanium alloys.
\newblock {\em Journal of Materials Processing Technology}, 133(1-2):84--89,
  2003.

\bibitem{gollapudi2011microstructure}
Srikant Gollapudi, R~Sarkar, U~Chinta Babu, R~Sankarasubramanian, TK~Nandy, and
  AK~Gogia.
\newblock Microstructure and mechanical properties of a copper containing three
  phase titanium alloy.
\newblock {\em Materials Science and Engineering: A}, 528(22-23):6794--6803,
  2011.

\bibitem{tang2015tensile}
Zhi Tang, Oleg~N Senkov, Chad~M Parish, Chuan Zhang, Fan Zhang, Louis~J
  Santodonato, Gongyao Wang, Guangfeng Zhao, Fuqian Yang, and Peter~K Liaw.
\newblock Tensile ductility of an alcocrfeni multi-phase high-entropy alloy
  through hot isostatic pressing (hip) and homogenization.
\newblock {\em Materials Science and Engineering: A}, 647:229--240, 2015.

\bibitem{liu2019fatigue}
Kaimiao Liu, Mageshwari Komarasamy, Bharat Gwalani, Shivakant Shukla, and
  Rajiv~S Mishra.
\newblock Fatigue behavior of ultrafine grained triplex al0. 3cocrfeni high
  entropy alloy.
\newblock {\em Scripta Materialia}, 158:116--120, 2019.

\bibitem{jang1995effect}
Byung-Koog Jang, Manabu Enoki, Teruo Kishi, and Hee-Kap Oh.
\newblock Effect of second phase on mechanical properties and toughening of
  al2o3 based ceramic composites.
\newblock {\em Composites Engineering}, 5(10-11):1275--1286, 1995.

\bibitem{lutz1991k}
Ekkehard~H Lutz, Nils Claussen, and Michael~V Swain.
\newblock K r-curve behavior of duplex ceramics.
\newblock {\em Journal of the American Ceramic Society}, 74(1):11--18, 1991.

\bibitem{neuman2017high}
Eric~W Neuman, Gregory~E Hilmas, and William~G Fahrenholtz.
\newblock A high strength alumina-silicon carbide-boron carbide triplex
  ceramic.
\newblock {\em Ceramics International}, 43(10):7958--7962, 2017.

\bibitem{feng2020effect}
Lun Feng, William~G Fahrenholtz, and Gregory~E Hilmas.
\newblock Effect of zrb2 content on the densification, microstructure, and
  mechanical properties of zrc-sic ceramics.
\newblock {\em Journal of the European Ceramic Society}, 40(2):220--225, 2020.

\bibitem{do2008microstructure}
AM~Do~Nascimento, V~Ocel{\'\i}k, MCF Ierardi, and J~Th~M De~Hosson.
\newblock Microstructure of reaction zone in wcp/duplex stainless steels matrix
  composites processing by laser melt injection.
\newblock {\em Surface and Coatings Technology}, 202(10):2113--2120, 2008.

\bibitem{sternitzke1997structural}
Martin Sternitzke.
\newblock Structural ceramic nanocomposites.
\newblock {\em Journal of the European Ceramic Society}, 17(9):1061--1082,
  1997.

\bibitem{fan1997computer}
Danan Fan and Long-Qing Chen.
\newblock Computer simulation of grain growth and ostwald ripening in
  alumina—zirconia two-phase composites.
\newblock {\em Journal of the American Ceramic Society}, 80(7):1773--1780,
  1997.

\bibitem{yu2021high}
Yongdong Yu, Fengyu Lin, Yongting Zheng, Wanjun Yu, Xudong Liu, YuChen Yuan,
  Hang Yin, and Xiaodong He.
\newblock High-density nanoprecipitation mechanism and microstructure evolution
  of high-performance al2o3/zro2 nanocomposite ceramics.
\newblock {\em Journal of the European Ceramic Society}, 41(10):5269--5279,
  2021.

\bibitem{cahn1991stability}
JW~Cahn.
\newblock Stability, microstructural evolution, grain growth, and coarsening in
  a two-dimensional two-phase microstructure.
\newblock {\em Acta metallurgica et materialia}, 39(10):2189--2199, 1991.

\bibitem{fan1997diffusion}
Danan Fan and Long-Qing Chen.
\newblock Diffusion-controlled grain growth in two-phase solids.
\newblock {\em Acta materialia}, 45(8):3297--3310, 1997.

\bibitem{guo2009microstructural}
Wei-Ming Guo, Guo-Jun Zhang, and Pei-Ling Wang.
\newblock Microstructural evolution and grain growth kinetics in zrb2--sic
  composites during heat treatment.
\newblock {\em Journal of the American Ceramic Society}, 92(11):2780--2783,
  2009.

\bibitem{liu2015synergetic}
Hu-Lin Liu, Guo-Jun Zhang, Ji-Xuan Liu, and Houzheng Wu.
\newblock Synergetic roles of zrc and sic in ternary zrb2--sic--zrc ceramics.
\newblock {\em Journal of the European Ceramic Society}, 35(16):4389--4397,
  2015.

\bibitem{lei2017phase}
Yinkai Lei, Tian-Le Cheng, and You-Hai Wen.
\newblock Phase field modeling of microstructure evolution and concomitant
  effective conductivity change in solid oxide fuel cell electrodes.
\newblock {\em Journal of Power Sources}, 345:275--289, 2017.

\bibitem{saito1992monte}
Yoshiyuki Saito and Masato Enomoto.
\newblock Monte carlo simulation of grain growth.
\newblock {\em ISIJ international}, 32(3):267--274, 1992.

\bibitem{kawasaki1989vertex}
Kyozi Kawasaki, Tatsuzo Nagai, and Katsuya Nakashima.
\newblock Vertex models for two-dimensional grain growth.
\newblock {\em Philosophical Magazine B}, 60(3):399--421, 1989.

\bibitem{anderson1984computer}
MP~Anderson, DJ~Srolovitz, GS~Grest, and PS~Sahni.
\newblock Computer simulation of grain growth—i. kinetics.
\newblock {\em Acta metallurgica}, 32(5):783--791, 1984.

\bibitem{krill2002computer}
CE~Krill~Iii and L-Q Chen.
\newblock Computer simulation of 3-d grain growth using a phase-field model.
\newblock {\em Acta materialia}, 50(12):3059--3075, 2002.

\bibitem{perumal2017phase}
Ramanathan Perumal, PG~Kubendran Amos, Michael Selzer, and Britta Nestler.
\newblock Phase-field study on the formation of first-neighbour topological
  clusters during the isotropic grain growth.
\newblock {\em Computational Materials Science}, 140:209--223, 2017.

\bibitem{mckenna2014grain}
IM~McKenna, SO~Poulsen, Erik~Mejdal Lauridsen, W~Ludwig, and Peter~W Voorhees.
\newblock Grain growth in four dimensions: A comparison between simulation and
  experiment.
\newblock {\em Acta materialia}, 78:125--134, 2014.

\bibitem{ravash2014three}
Hamed Ravash, Jef Vleugels, and Nele Moelans.
\newblock Three-dimensional phase-field simulation of microstructural evolution
  in three-phase materials with different diffusivities.
\newblock {\em Journal of materials science}, 49(20):7066--7072, 2014.

\bibitem{ravash2017three2}
Hamed Ravash, Jef Vleugels, and Nele Moelans.
\newblock Three-dimensional phase-field simulation of microstructural evolution
  in three-phase materials with different interfacial energies and different
  diffusivities.
\newblock {\em Journal of materials science}, 52(24):13852--13867, 2017.

\bibitem{yadav2016effect}
Vishal Yadav, Liesbeth Vanherpe, and Nele Moelans.
\newblock Effect of volume fractions on microstructure evolution in isotropic
  volume-conserved two-phase alloys: A phase-field study.
\newblock {\em Computational Materials Science}, 125:297--308, 2016.

\bibitem{ohnuma1999computer}
Ikuo Ohnuma, Kiyohito Ishida, and Taiji Nishizawa.
\newblock Computer simulation of grain growth in dual-phase structures.
\newblock {\em Philosophical Magazine A}, 79(5):1131--1144, 1999.

\bibitem{fan1997topological}
Danan Fan and Long-Qing Chen.
\newblock Topological evolution during coupled grain growth and ostwald
  ripening in volume-conserved 2-d two-phase polycrystals.
\newblock {\em Acta materialia}, 45(10):4145--4154, 1997.

\bibitem{perumal2018phase}
Ramanathan Perumal, PG~Kubendran Amos, Michael Selzer, and Britta Nestler.
\newblock Phase-field study of the transient phenomena induced by
  ‘abnormally’large grains during 2-dimensional isotropic grain growth.
\newblock {\em Computational Materials Science}, 147:227--237, 2018.

\bibitem{amos2019understanding}
PG~Amos.
\newblock Understanding the volume-diffusion governed shape-instabilities in
  metallic systems.
\newblock {\em arXiv preprint arXiv:1906.10404}, 2019.

\bibitem{amos2020grand}
PG~Kubendran Amos and Britta Nestler.
\newblock Grand-potential based phase-field model for systems with interstitial
  sites.
\newblock {\em Scientific reports}, 10(1):1--22, 2020.

\bibitem{amos2020distinguishing}
PG~Kubendran Amos and Britta Nestler.
\newblock Distinguishing interstitial and substitutional diffusion in
  grand-potential based phase-field model.
\newblock {\em Materialia}, 12:100820, 2020.

\bibitem{perumal2019concurrent}
Ramanathan Perumal, Michael Selzer, and Britta Nestler.
\newblock Concurrent grain growth and coarsening of two-phase microstructures;
  large scale phase-field study.
\newblock {\em Computational Materials Science}, 159:160--176, 2019.

\bibitem{perumal2020quadrijunctions}
Ramanathan Perumal, PG~Kubendran Amos, Michael Selzer, and Britta Nestler.
\newblock Quadrijunctions-stunted grain growth in duplex microstructure: a
  multiphase-field analysis.
\newblock {\em Scripta Materialia}, 182:16--20, 2020.

\bibitem{amos2020multiphase}
PG~Kubendran Amos, Ramanathan Perumal, Michael Selzer, and Britta Nestler.
\newblock Multiphase-field modelling of concurrent grain growth and coarsening
  in complex multicomponent systems.
\newblock {\em Journal of Materials Science \& Technology}, 45:215--229, 2020.

\bibitem{hoffrogge2021multiphase}
Paul~W Hoffrogge, Arnab Mukherjee, ES~Nani, PG~Kubendran Amos, Fei Wang, Daniel
  Schneider, and Britta Nestler.
\newblock Multiphase-field model for surface diffusion and attachment kinetics
  in the grand-potential framework.
\newblock {\em Physical Review E}, 103(3):033307, 2021.

\bibitem{provatas2011phase}
Nikolas Provatas and Ken Elder.
\newblock {\em Phase-field methods in materials science and engineering}.
\newblock John Wiley \& Sons, 2011.

\bibitem{tschukin2017concepts}
Oleg Tschukin, Alexander Silberzahn, Michael Selzer, Prince~GK Amos, Daniel
  Schneider, and Britta Nestler.
\newblock Concepts of modeling surface energy anisotropy in phase-field
  approaches.
\newblock {\em Geothermal Energy}, 5(1):1--21, 2017.

\bibitem{amos2018phase}
PG~Kubendran Amos, LT~Mushongera, and Britta Nestler.
\newblock Phase-field analysis of volume-diffusion controlled
  shape-instabilities in metallic systems-i: 2-dimensional plate-like
  structures.
\newblock {\em Computational Materials Science}, 144:363--373, 2018.

\bibitem{amos2018globularization}
PG~Kubendran Amos, Ephraim Schoof, Daniel Schneider, and Britta Nestler.
\newblock On the globularization of the shapes associated with
  alpha-precipitate of two phase titanium alloys: Insights from phase-field
  simulations.
\newblock {\em Acta Materialia}, 159:51--64, 2018.

\bibitem{amos2020limitations}
PG~Kubendran Amos, Ephraim Schoof, Jay Santoki, Daniel Schneider, and Britta
  Nestler.
\newblock Limitations of preserving volume in allen-cahn framework for
  microstructural analysis.
\newblock {\em Computational Materials Science}, 173:109388, 2020.

\bibitem{plapp2011unified}
Mathis Plapp.
\newblock Unified derivation of phase-field models for alloy solidification
  from a grand-potential functional.
\newblock {\em Physical Review E}, 84(3):031601, 2011.

\bibitem{fan1998numerical}
Danan Fan, Long-Qing Chen, and Shao-Ping~P Chen.
\newblock Numerical simulation of zener pinning with growing second-phase
  particles.
\newblock {\em Journal of the American Ceramic Society}, 81(3):526--532, 1998.

\bibitem{poulsen2013three}
Stefan~Othmar Poulsen, PW~Voorhees, and Erik~Mejdal Lauridsen.
\newblock Three-dimensional simulations of microstructural evolution in
  polycrystalline dual-phase materials with constant volume fractions.
\newblock {\em Acta materialia}, 61(4):1220--1228, 2013.

\bibitem{mittnacht2021morphological}
Tobias Mittnacht, PG~Kubendran Amos, Daniel Schneider, and Britta Nestler.
\newblock Morphological stability of three-dimensional cementite rods in
  polycrystalline system: a phase-field analysis.
\newblock {\em Journal of Materials Science \& Technology}, 77:252--268, 2021.

\bibitem{ritasalo2013microstructural}
R~Ritasalo, ME~Cura, XW~Liu, Y~Ge, Topi Kosonen, Ulla Kanerva, O~S{\"o}derberg,
  and SP~Hannula.
\newblock Microstructural and mechanical characteristics of cu--cu2o composites
  compacted with pulsed electric current sintering and hot isostatic pressing.
\newblock {\em Composites Part A: Applied Science and Manufacturing},
  45:61--69, 2013.

\bibitem{praveen2016exceptional}
S~Praveen, Joysurya Basu, Sanjay Kashyap, and Ravi~Sankar Kottada.
\newblock Exceptional resistance to grain growth in nanocrystalline cocrfeni
  high entropy alloy at high homologous temperatures.
\newblock {\em Journal of Alloys and Compounds}, 662:361--367, 2016.

\end{thebibliography}
\end{document}